\documentclass[lettersize,journal]{IEEEtran}
\usepackage{amsmath,amsfonts}
\usepackage{array}
\usepackage{textcomp}
\usepackage{stfloats}
\usepackage{url}
\usepackage{verbatim}
\usepackage{cite}
\usepackage{graphicx}  
\usepackage[caption=false,font=footnotesize]{subfig}
\usepackage{tabularx}
\usepackage{relsize}
\usepackage{multirow}
\usepackage{booktabs}

\usepackage[ruled,vlined,linesnumbered]{algorithm2e}

\begin{document}

\title{Optimizing Winograd Convolution on ARMv8 Processors}

\author{%
    Haoyuan~Gui,
    Xiaoyu~Zhang,
    Chong~Zhang,
    and~Huiyuan~Li
    \thanks{Haoyuan Gui, Xiaoyu Zhang and Chong Zhang are with the Institute of Software, Chinese Academy of Sciences,
            Beijing 100190, China, and also with the University of Chinese Academy of Sciences,
            Beijing 100049, China (e-mail: guihaoyuan123@icloud.com, zhangxy420@foxmail.com, zhangchong2020@iscas.ac.cn).}%
    \thanks{Huiyuan Li is with the Institute of Software, Chinese Academy of Sciences,
            Beijing 100190, China, and also with the State Key Laboratory of Computer Science,
            Chinese Academy of Sciences, Beijing 100190, China (e-mail: huiyuan@iscas.ac.cn).}%
}

\maketitle

\begin{abstract}
  As Convolutional Neural Networks (CNNs) gain increasing prominence in deep learning, algorithms such as Winograd convolution have been introduced to boost computational efficiency. 
  However, existing Winograd implementations often face challenges, including significant transformation overhead, suboptimal computation efficiency, 
  and reduced parallel performance in certain layers. In this work, we present an ARMv8-specific execution framework for Winograd convolution based on a cache-resident, block-wise schedule. 
  By maintaining consecutive memory access where possible and packing data in a customized z-shaped layout, our approach fully exploits a meticulously optimized GEMM micro-kernel that employs a ping-pong technique. 
  Furthermore, we introduce a multidimensional parallelization strategy that selects a thread decomposition across the batch, tile, input-channel, and output-channel loops according to layer shape and coordinates the resulting thread groups with the fused block-wise execution schedule.
  To further boost performance, we manually optimize each micro-kernel in AArch64 assembly, 
  carefully tuning blocking parameters for the ARMv8 architecture to minimize transformation overhead and maximize computational throughput. 
  Experimental results demonstrate that our method achieves up to $4.74\times$, $4.10\times$, $4.72\times$, and $10.57\times$ speedups over NCNN, NNPACK, FastConv, and ACL, respectively, 
  on the Kunpeng 920 platform when using multiple threads, with corresponding gains of
  $5.46\times$, $2.81\times$, $6.11\times$, and $8.05\times$ on the AWS Graviton2, 
  and $3.32\times$, $3.68\times$, $5.84\times$, and $9.28\times$ on the Phytium 2000+.
\end{abstract}


\section{Introduction}
\IEEEPARstart{C}{onvolutional} Neural Networks (CNNs) have achieved great advancements and demonstrated exceptional performance across various fields of artificial intelligence \cite{ref1, ref2, ref3, ref4, ref5}.
Despite their impressive accuracy and versatility, CNNs require substantial computational resources, primarily due to the computational intensive operations within their convolutional layers.
As CNN models continue to grow in complexity and depth, the need for efficient optimization techniques becomes increasingly critical.

Lavin and Gray \cite{ref6} proposed leveraging the Winograd minimal filtering algorithm \cite{ref7} to mitigate the computational complexity inherent in convolution operations. 
This efficient algorithm divides the input of convolutional networks into numerous tiles, 
transforms these tiles and the corresponding filters into the Winograd domain, performs element-wise multiplications, and subsequently transforms the results back into the spatial domain.
This approach reduces the number of arithmetic operations required, 
thereby enhancing the computational efficiency of convolutional neural networks.
Consequently, it has been incorporated into contemporary deep learning libraries. 
For instance, CuDNN \cite{ref8} and MIOPEN \cite{ref9} have implemented Winograd-based convolution for Nvidia GPUs and AMD GPUs, respectively. 
Similarly, OneDNN \cite{ref10} and FALCON \cite{ref11} have implemented this approach for x86 CPUs. 
Additionally, libraries such as NCNN \cite{ref12}, NNPACK \cite{ref13}, FastConv \cite{ref14} and the ARM Compute Library (ACL) \cite{ref15} have been designed specifically for ARM CPUs. 
The ARM architecture predominates in mobile computing and is also
increasingly used in high-performance computing (HPC) systems for
convolutional training and inference.

On ARMv8 processors, however, the multiplication savings of Winograd
convolution do not directly translate into proportional runtime gains.
Performance also depends on data movement, SIMD utilization, and thread
scheduling. We therefore focus on four coupled components of an ARMv8
Winograd implementation: the runtime schedule, the micro-kernels, the
intermediate-data layout, and the parallel strategy.

  The main contributions of this work are summarized as follows:
  \begin{itemize}
    \item We design a fused block-wise execution schedule for the runtime stages of input transformation, GEMM, and output transformation. A performance model guides the choice of blocking parameters under L1 and L2 cache-capacity constraints. Compared with the stage-wise non-fused schedule used in our ablation study, the resulting tile-block organization shortens the lifetime of intermediate data and reduces the active working-set size.

    \item We co-design hand-tuned AArch64 assembly micro-kernels for the transformation and GEMM stages with a custom z-shaped intermediate-data layout. Two register-aware transformation schemes improve data reuse and reduce load and fused multiply-add (FMA) instruction counts. The transformation kernels write their outputs directly into this layout in the order consumed by GEMM, which avoids a separate data-repacking step. The GEMM micro-kernels share a register-blocked ping-pong design that uses the available SIMD registers to overlap data loading with computation. Both sets of micro-kernels use software prefetching.

    \item We develop a multidimensional parallelization strategy that distributes threads across the batch, tile, input-channel, and output-channel dimensions according to layer shape. The resulting thread decomposition is coordinated with the block-wise execution schedule to improve load balance and parallel efficiency across convolutional layers of different sizes.
  \end{itemize}
  By applying these optimizations, our method achieves up to $4.74\times$, $4.10\times$, $4.72\times$, and $10.57\times$ speedups over NCNN, NNPACK, FastConv, and ACL respectively using multiple threads on the Kunpeng 920 platform, 
    $5.46\times$, $2.81\times$, $6.11\times$, and $8.05\times$ on the AWS Graviton2, 
    and $3.32\times$, $3.68\times$, $5.84\times$, and $9.28\times$ on the Phytium 2000+ platform.

  The remainder of this paper is organized as follows: Section \ref{sec:background} provides the background of Winograd convolution. 
  Section \ref{sec:systemdesign} details our system design framework, including the kernels for transformation and GEMM, as well as the design of the parallel strategy. 
  Section \ref{sec:experiments} evaluates the performance of our approach in comparison to NCNN, NNPACK, FastConv and ACL. 
  Section \ref{sec:relatedwork} discusses related work on Winograd convolution, and Section \ref{sec:conclusion} concludes the paper and discusses future research directions.

\section{BACKGROUND}
\label{sec:background}
\subsection{Convolutional Neural Networks}

In a convolutional neural network layer, a filter tensor \(F\) with the dimensions
 \(K \times C \times R \times S \) is applied to an input tensor  \(D\) 
 which is  shaped \(N \times C \times H \times W \). In this context, \(K\) and \(C\) represent 
 the number of output and input channels, respectively. The dimensions $R$ and $S$ (resp. $H$ and $W$) 
 correspond to  the height and width of the filter (resp. input). Finally, \(N\) denotes the batch size of the input.
 For the following equation, each element of \(F\) and \(D\) is denoted as \(F_{k,c,u,v}\) and \(D_{b,c,i,j}\), respectively.
 The corresponding output tensor \(O\) for the \(b-th\) batch and \(k-th\) output channels can be calculated by  Equation \eqref{eq:1},
 \begin{equation}
  O_{b,k,i,j} = \sum_{c=1}^{C} \sum_{u=1}^{R} \sum_{v=1}^{S} D_{b,c,i+u,j+v} \cdot F_{k,c,u,v}.
  \label{eq:1}
 \end{equation}
 Thus, it can also be rewritten as 
 \begin{equation}
  O_{b,k} = \sum_{c=1}^{C} D_{b,c} * F_{k,c},
  \label{eq:2}
 \end{equation}  

 where * denotes convolution operation.

 \subsection{Winograd Convolution}
Winograd convolution \cite{ref6} leverages the Winograd minimal filtering algorithm \cite{ref7} to reduce arithmetic complexity of the convolution process.
The notation \(F(m,r)\) represents using an \(r\)-tap FIR filter to compute \(m\) outputs, where the input size is \(m+r-1\). This method can reduce the number of multiplications by a factor of \((m \times r)/(m+r-1)\) 
 compared to the direct convolution. For \(n\)-dimensional, this notation extends to \(F(m_1 \times m_2 \times ... \times m_n, r_1 \times r_2 \times ... \times r_n)\), 
where the length of output and filter in the \(i\)-th dimension are \(m_i\) and \(r_i\), respectively.

To illustrate, we use 2-D convolution as an example. Winograd convolution employs overlap-add (OLA) method, splitting \(D_{b,c}\) with dimensions \(H \times W\) into \([(H - r_1 + 1)/m_1] \times [(W - r_2 + 1)/m_2]\) tiles. Each tile 
contains \((m_1 + r_1 - 1) \times (m_2 + r_2 - 1)\) elements, with $r_i -1$ elements overlapping with neighbouring tiles in each dimension.

The key idea of Winograd convolution is to transform the input tensor and filter tensor from spatial domain into Winograd domain before computation, thereby to reduce the number of multiplications. 
For instance, the \(F(2 \times 2, 3 \times 3)\) configuration achieves a theoretical speedup of  $2.25\times$, \(F(4 \times 4, 3 \times 3)\) attains  $4\times$ and for \(F(6 \times 6, 3 \times 3)\), this value can reach $5.0625\times$. After computing, the output tensor \( \hat{O}\) in Winograd domain will be transformed back. 
Assume that $\hat{i}$ and $\hat{j}$ are the row and column tile coordinates of the input, so that we have $d = D_{b, c, \hat{i}, \hat{j}}$ and $g = F_{k,c}$. 
The corresponding output $O_{b, k, \hat{i}, \hat{j}}$ can be calculated by Winograd convolution via Equation \eqref{eq:3} 
\begin{equation}
  \begin{split}
  O_{b,k,\hat{i},\hat{j}}
    &= \sum_{c=1}^{C}
       A^{T}\!\left[\left(GgG^{T}\right)\odot\left(B^{T}dB\right)\right]A\\
    &= A^{T}\!\left[\sum_{c=1}^{C}
       \left(U_{k,c}\odot V_{b,c,\hat{i},\hat{j}}\right)\right]A.
  \end{split}
  \label{eq:3}
\end{equation}
where $\odot$ symbolizes element-wise multiplication, and $B$, $G$, $A$ are transformation matrices for input, filter and output respectively. 
For simplicity, the coordinates $(b,\hat{i},\hat{j})$ can be collapsed down to a single dimension $\xi$. 
We use \((x,y)\) to denote the coordinates of elements involved in the element-wise multiplication, satisfying \( 1 \leq x, y \leq m + r - 1 \).
By performing this transformation, Equation \eqref{eq:3}
can be converted into its new form as follows:
\begin{equation}
  \hat{O}^{(x,y)}_{k,\xi} =  \sum_{c=1}^{C} U^{(x,y)}_{k,c}V^{(x,y)}_{c,\xi}.
  \label{eq:4}
\end{equation}

The progress of Equation \eqref{eq:4} can be viewed as matrix multiplication, a Level-3 Basic Linear Algebra Subprograms (BLAS) operation with higher arithmetic intensity compared to element-wise multiplication, which is a Level-1 BLAS operation. 
The transformation matrices used in Winograd convolution can be generated using the Chinese Remainder Theorem (CRT) \cite{ref7}.
To conserve space, we only provide the input transformation matrices \(B^T\) for \(F(2 \times 2, 3 \times 3)\) and \(F(6 \times 6, 3 \times 3)\) below:

 \begin{equation}
  \begin{aligned}
  B^{T}_{2,3} &=
  \begin{bmatrix}
    1 & 0 & -1 & 0 \\
    0 & 1 &  1 & 0 \\
    0 & -1 & 1 & 0 \\
    0 & -1 & 0 & 1
  \end{bmatrix}, \\[6pt]   
  B^{T}_{6,3} &=
  \begin{bmatrix}
    1 & 0 & -\frac{21}{4} & 0 &  \frac{21}{4} & 0 & -1 & 0 \\
    0 & 1 &  1 & -\frac{17}{4} &  \frac{17}{4} & 1 &  1 & 0 \\
    0 & -1 & 1 &  \frac{17}{4} & -\frac{17}{4} & -1 & 1 & 0 \\
    0 & -\frac{1}{2} &  \frac{1}{4} & -\frac{5}{2} & -\frac{5}{4} & 2 & 1 & 0 \\
    0 & -\frac{1}{2} &  \frac{1}{4} &  \frac{5}{2} & -\frac{5}{4} & -2 & 1 & 0 \\
    0 & 2 & 4 & -\frac{5}{2} & -5 &  \frac{1}{2} & 1 & 0 \\
    0 & -2 & 4 &  \frac{5}{2} & -5 & -\frac{1}{2} & 1 & 0 \\
    0 & -1 & 0 &  \frac{21}{4} & 0 & -\frac{21}{4} & 0 & 1
  \end{bmatrix}.
  \end{aligned}
  \label{eq:5}
  \end{equation}

  \section{SYSTEM DESIGN}
  \label{sec:systemdesign}
\begin{figure*}[t]
  \centering
  \includegraphics[scale=1]{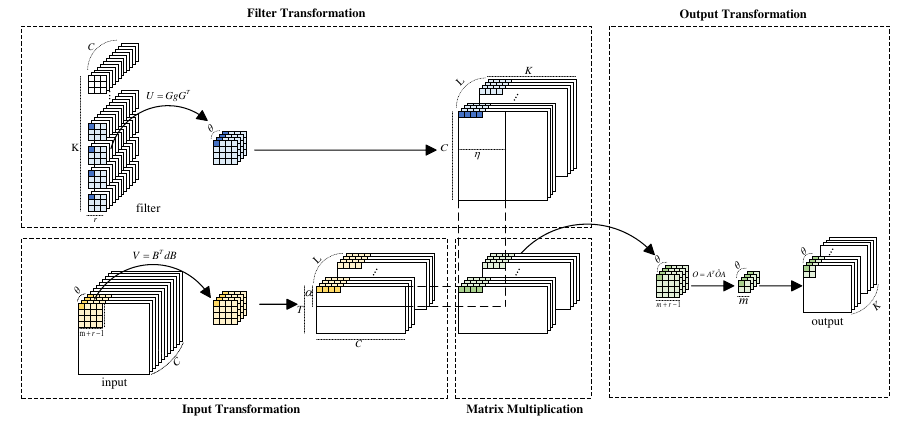}
  \caption{Overview of the procedure for processing one tile of input using our method, 
  which includes the three stages of Winograd convolution: Input/Filter Transformation, Matrix Multiplication, 
  and Output Transformation. The yellow, blue, and green rectangles represent the data of the input, filter, and output, respectively. 
  The highlighted sections of each color indicate the data loaded into the same vector register, which will be processed simultaneously. 
  After transforming the input and filter, the data is packed into a layout that is friendly to GEMM operations, ensuring consecutive memory access during computation. 
  The results of the GEMM are then transformed back to the spatial domain and stored in the final output.}
  \label{fig:fig_1}
\end{figure*}

  Our ARMv8 framework uses a block-wise runtime schedule in which input transformation, GEMM, and output transformation are coordinated at tile-block granularity. Filter transformation is performed before the main tile loop and can be precomputed for inference. Hand-tuned AArch64 assembly micro-kernels implement the transformation and GEMM stages. The transformation kernels write their outputs directly to GEMM-ready layouts. Blocking parameters are selected under L1 and L2 cache-capacity constraints, while the parallel strategy distributes threads across the batch, tile, input-channel, and output-channel loops according to layer shape. Fig.~\ref{fig:fig_1} shows the overall data flow. The following notation is used:
  \begin{itemize}
   \item \textbf{$\theta$}: The number of floating-point numbers that can be stored in a vector register. For ARMv8, the bit-width of a vector register is 128 bits, so $\theta$ is 4 for FP32.
   \item \textbf{$L$}: We assume $L = (m+r-1) \times (m+r-1)$, which represents the number of elements in one Winograd tile and the batch size for GEMM.
   \item \textbf{$T$}: The number of tiles.
   \item \textbf{($\alpha, \eta$)}: The parameters of our GEMM micro-kernel. For each computation, it multiplies matrix $V$ with $\alpha$ rows by matrix $U$ with $\eta$ columns, producing a matrix $\hat{O}$ of shape $(\alpha, \eta)$.
 \end{itemize}

\begin{algorithm}[t]
  \caption{Winograd Implementation with GEMM for a single batch}
  \label{alg:alg1}
  \SetAlgoNoEnd
  \SetAlgoLined
  \KwIn{Input[C][H][W], Filter[K][C][R][S]}
  \KwOut{Output[K][P][Q]}
  \tcp{Allocate temporary arrays:}
  $TransInOut \leftarrow \text{new array}[L \times T_{blk} \times C ]$\;
  $GEMMOut \leftarrow \text{new array}[L \times T_{blk} \times K_{blk}]$\;
  $FilterOut \leftarrow \text{new array}[L \times C \times K]$;

  \For{$bk = 0$ \KwTo $K$ \KwSty{step}  $K_{blk}$}{
    \For{$bc = 0$ \KwTo $C$ \KwSty{step}  $C_{blk}$}{
      \KwSty{Filter Transform Kernel ;} 
      \tcp{Transform and pack $C_{blk} \times K_{blk}$ tiles into Winograd domain, storing results in $FilterOut[0 : L, bc : bc + C_{blk}, bk : bk + K_{blk}]$.}
    }
  }
  \For{$bt = 0$ \KwTo $T - T\%T_{blk}$ \KwSty{step}  $T_{blk}$}{
    \For{$bc = 0$ \KwTo $C$ \KwSty{step}  $C_{blk}$}
    {
      \tcp{Input Transform Kernel}
      \For{$ic = 0$ \KwTo $C_{blk}$ \KwSty{step}  $C_{T}$}{
        \KwSty{Input Transform micro-kernel ;} 
        \tcp{Transform and pack $C_{T} \times T_{blk}$ tiles into Winograd domain, storing results in $TransInOut[0 : L, 0 : T_{blk}, bc + ic: bc + ic + C_{T}]$.}
      }
    }
    \For{$bk = 0$ \KwTo $K$ \KwSty{step}  $K_{blk}$}{
      \For{$i = 0$ \KwTo $L$ \KwSty{step}  $1$}{
        \For{$bc = 0$ \KwTo $C$ \KwSty{step}  $C_{blk}$}{
          \KwSty{GEMM Kernel ;} 
          \tcp{Multiply and accumulate the result to $GEMMOut[i, 0 : T_{blk}, 0 : K_{blk}]$}
        }
      }
      \KwSty{Output Transform Kernel ;}
      \tcp{Transform $GEMMOut[0 : L, 0 : T_{blk}, bk : bk + K_{blk}]$ back to the spatial domain and store in Output}
    }
  }
  Process the remaining $T\%T_{blk}$ tiles. The framework follows the same procedure as described above, but with kernels utilizing different values of $\alpha$ and $\eta$. 
\end{algorithm}

Algorithm~\ref{alg:alg1} specifies the loop structure of the block-wise runtime schedule. Here, $T_{blk}$, $C_{blk}$, and $K_{blk}$ denote the block sizes for the tile, input-channel, and output-channel dimensions, respectively; they correspond to the $M$, $K$, and $N$ dimensions of GEMM. $C_T$ denotes the number of input channels processed by each input transformation micro-kernel.
Compared with a standalone GEMM loop nest such as GotoBLAS~\cite{ref19}, this schedule must also account for when transformed inputs are produced and when GEMM outputs are consumed by the output transformation.
Here, fusion refers to schedule-level fusion rather than a single
monolithic kernel. The three stages are implemented by separate
micro-kernels but are invoked within the same tile-block loop, allowing
transformed inputs and GEMM outputs to be consumed before execution
advances to the next tile block.

The input and filter transformation kernels write their packed results
directly to \textit{TransInOut} and \textit{FilterOut}, respectively.
The two $z$-shaped layouts follow the operand-access order of the GEMM
micro-kernel, which accesses the packed operands consecutively as it
advances along the $C$ dimension. During each tile-block iteration,
GEMM stores the resulting $L \times T_{blk} \times K_{blk}$ block in
\textit{GEMMOut}. Once the block is complete, the
output transformation kernel reads it and writes the corresponding
spatial-domain output.

Hand-tuned AArch64 assembly gives the transformation and GEMM
micro-kernels explicit control over SIMD register allocation and
software prefetching, allowing data movement to be coordinated with
computation.

Packed layouts must match the block boundaries and GEMM access order
so that transformation results can be consumed directly. Block-wise
execution reuses intermediate buffers across tile blocks, while L1
and L2 capacity constraints limit the block dimensions. Increasing
$T_{blk}$ improves filter reuse but reduces the number of independent
tile blocks available for parallel execution.

When tile parallelism is insufficient, threads within a group transform
different input channels of the same tile block in parallel. After
synchronization, they share the transformed input and process different
output-channel blocks through GEMM and output transformation. This adds
parallelism while retaining buffer reuse, at the cost of separate GEMM
output buffers and group synchronization. Table~\ref{tab:scheduling_ablation}
evaluates the complementary benefits of block-wise scheduling and thread
decomposition.

\subsection{Input and filter transformation}
\label{subsec: input_filter_transformation}
As neural networks deepen, the spatial dimensions of the feature maps typically decrease, often due to pooling operations\cite{ref20}, while the number of channels simultaneously increases.
This change in dimensions and channel size is a common characteristic of deep neural networks, enabling them to capture more complex features. 
Table~\ref{tab:table1} lists the parameters of layers for two mainstream convolutional neural networks.
We assume that $t_i$ and $t_f$ respectively denote the number of transform operation for input and filter,
where $t_i$ is proportional to  $T \times C = \frac{(H - r + 1)(W - r + 1)C}{m^2} \propto H \times W \times C$ and $t_f$ is proportional to $C \times K $. 
Given the trends of $H \& W$, $C$ and $K$ as shown in table~\ref{tab:table1},
the transformation time for the input will decrease, while that for the filter will increase. 
Consequently, these transformations become bottlenecks in different parts of the neural network: the input transformations in the shallower layers and the filter transformations in the deeper layers.
Therefore, it is crucial to optimize both input and filter transformations to enhance the overall performance of the Winograd convolution.

\subsubsection{Transformation}
We implement three Winograd variants: $F(2\times2,3\times3)$,
$F(4\times4,3\times3)$, and $F(6\times6,3\times3)$. Their
transformation kernels use two strategies based on SIMD-register
demand. For $F(2\times2,3\times3)$, $2L=32$, allowing the input tile
and its transformed result to be held simultaneously in the 32 SIMD
registers. For $F(4\times4,3\times3)$ and $F(6\times6,3\times3)$,
$2L>32$, so the kernels process the transformation one row at a time.
These two variants use the same row-wise procedure; we describe this
strategy using $F(6\times6,3\times3)$. The transformation matrices are
obtained from \textit{wincnn}~\cite{ref21}. In both strategies, each
vector register holds $\theta$ values at the same coordinate $(x,y)$
from $\theta$ channels, as defined in Equation~\eqref{eq:4}. The input
and filter transformations share the same basic structure, so the
following discussion focuses on input transformation.

\begin{figure}[h]
  \centering
  \includegraphics[scale=0.8]{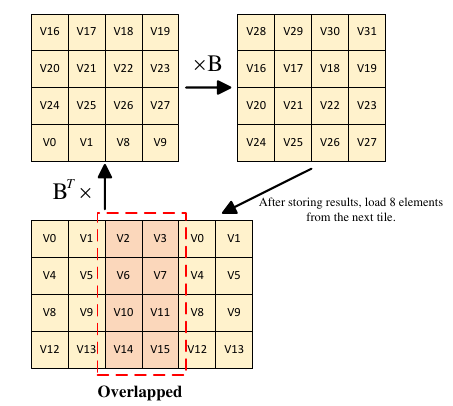}
\caption{Processing flow and register organization in the
$F(2\times2,3\times3)$ input transformation micro-kernel. Each vector
register holds $C_T=\theta$ values at the same tile coordinate across
$\theta$ channels, and the labels denote vector-register indices. The
micro-kernel loads the first tile into $v0$ through $v15$, applies
$B^T$ and $B$, and stores the transformed tile in $v16$ through $v31$.
Registers released during the $B^T$ multiplication are reused to hold
intermediate values.}
  \label{fig:fig_2}
\end{figure}

\textbf{F($2 \times 2$, $3 \times 3$):} The ARMv8 architecture provides 32 vector registers, each with a 128-bit width.
Under this configuration, each tile comprises $4\times4=16$ elements,
enabling simultaneous holding of 16 input elements and their corresponding 16 transformed results across $\theta=4$ channels within the available registers.
Due to the OLA method employed by the Winograd convolution and the row-major storage layout, 
adjacent tiles along the $W$ dimension share 8 elements, enabling register reuse.
Given the processing order ($W \rightarrow H \rightarrow C$), 
the initial pass along the $W$ dimension requires loads all 16 elements, 
while subsequent tiles require loading only 8 new elements each.
This approach can nearly halve the number of load instructions required. 
Details of the pipeline are illustrated in Fig.~\ref{fig:fig_2}, 
where each number denotes a vector register index.
Furthermore, this arrangement allows us to omit computations involving zero-valued elements in the transformation matrix $B_{2,3}$ (see Equation \eqref{eq:5}), 
and simplifies operations involving elements equal to one to mere addition or subtraction.
These optimizations markedly decrease computational overhead.

\textbf{F($6 \times 6$, $3 \times 3$):} In this configuration, each tile consists of $8 \times 8 = 64$ elements, exceeding the available vector registers on ARMv8 architectures. 
Consequently, only a subset of the tile elements can be processed at a time.
Our design strategy involves processing one row (comprising 8 elements) per iteration, simultaneously handling elements across $C_T = 2\theta$ channels to fully exploit register utilization.
Given the row-major data layout, 
we initially perform the multiplication $d \times B$, 
storing the intermediate results in a temporary array of dimensions $8\times8\times8$. 
Subsequently, this intermediate tensor is multiplied from the left by $B^T$. Exploiting the structural sparsity and redundancy inherent in the transformation matrix $B_{6,3}$, 
common computational factors can be extracted to significantly reduce arithmetic complexity. 
The computation corresponding to the $i$-th row follows Equation~\eqref{eq:7}, 
equivalently performing a left multiplication by the $i$-th column of $B^T_{6,3}$.

\begin{equation}
  \setlength{\arraycolsep}{4pt}  
  \begin{aligned}
    f_0 &= d^{(i,3)} + d^{(i,7)} - 4.25 \times d^{(i,4)} ,\\
    f_1 &= d^{(i,2)} - 4.25\times d^{(i,4)} +d^{(i,6)} ,\\
    f_2 &= 1.25 \times d^{(i,5)},\\
    f_3 &= 2.5 \times d^{(i,4)},\\
    f_4 &= 0.25 \times d^{(i,3)} - f_2 + d^{(i,7)},  \\
    f_5 &= 0.5 \times d^{(i,2)} - f_3 + 2\times d^{(i,6)},  \\
    f_6 &= 4 \times (d^{(i,3)} - f_2) +  d^{(i,7)}, \\
    f_7 &= 2 \times d^{(i,2)} - f_3 + 0.5\times d^{(i,6)},\\
  \end{aligned}
\end{equation}
\begin{equation}
  \begin{aligned}
  {tmp^i}^{\!T} &=
  \begin{bmatrix}
    d^{(i,1)} + 5.25 \times (d^{(i,5)}-d^{(i,3)}) - d^{(i,7)}\\
    f_0 + f_1\\
    f_0 - f_1\\
    f_4 + f_5\\
    f_4 - f_5\\
    f_6 + f_7\\
    f_6 - f_7\\
    5.25 \times (d^{(i,4)}-d^{(i,6)}) - d^{(i,2)} + d^{(i,8)}
  \end{bmatrix}.
  \end{aligned}
  \label{eq:7}
  \end{equation}

  Unlike input transformation, filter transformation involves both $C$ and $K$ dimensions,  
  introducing additional considerations for data loading into vector registers. 
  In the spatial domain, loading data to the same vector register across the $C$ dimension ensures consecutive memory access. 
  However, storing the transformed results becomes inefficient, 
  since $\eta$ (i.e., $K$) is the fastest-varying direction in the Winograd domain due to the GEMM-friendly data layout.
  To maximize data loading efficiency and enhance vector register utilization during storage, 
  the filter transformation loads $\theta$ channels along the $K$ dimension and processes them in the order of $\theta \rightarrow C \rightarrow \frac{K}{\theta}$.

\subsubsection{Data packing}
\begin{figure}[t]
  \centering
  \subfloat[]{\includegraphics[width=0.8\linewidth]{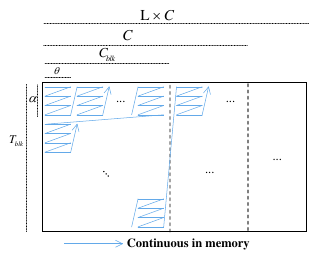}\label{fig:fig_3a}}\\
  \subfloat[]{\includegraphics[width=0.9\linewidth]{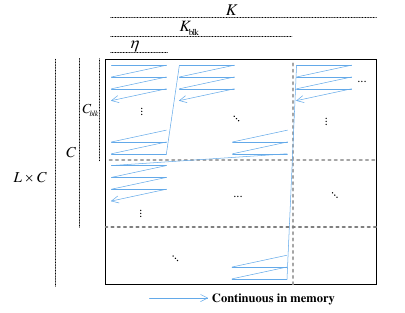}\label{fig:fig_3b}}
  \caption{Blocked $z$-shaped layouts of (a) \textit{TransInOut} and
  (b) \textit{FilterOut}. Dashed lines delimit data blocks, and arrows
  indicate the order in which elements are stored contiguously in memory.}
  \label{fig:fig_3}
\end{figure}

In order to ensure consecutive memory access for Matrix Multiplication, we store the transformed results in GEMM friendly format. The data layout of  \textit{TransInOut} and \textit{FilterOut} are [$L$][$C/C_{blk}$][$T_{blk}/\alpha$][$C_{blk}/\theta$][$\alpha$][$\theta$]
and [$K/K_{blk}$][$L$] [$C/C_{blk}$][$K_{blk}/\eta$][$C_{blk}$][$\eta$] as showed in Fig.~\ref{fig:fig_3a} and Fig.~\ref{fig:fig_3b}, respectively. Here, $\eta$ is a multiple of $\theta$ for vectorization load.
Since the memory access for spatial input spans the $C$ dimension and the transformed results are scattered across $L$ different memory blocks, ensuring consecutive memory access is challenging. Our goal is to mitigate this issue.
For F($2 \times 2$, $3 \times 3$), we reduce the number of load instructions by first looping in the $W$ direction, as previously mentioned. For F($6 \times 6$, $3 \times 3$), the processing order of our approach is  $C \rightarrow W \rightarrow H$,
ensuring consecutive memory access within each coordinate $(x,y)$ corresponding memory area.

\subsection{Matrix Multiplication}
The matrix-multiplication design has two components: GEMM micro-kernels
and cache blocking. Each micro-kernel computes an
$\alpha \times \eta$ output block from $\alpha$ rows of the blocked
transformed input matrix and $\eta$ columns of the blocked transformed
filter matrix. The micro-kernel design coordinates register allocation
and instruction scheduling to increase the computation-to-memory ratio
(CMR) and overlap operand loading with FMA execution, while two
configurations reduce edge overhead across different layer shapes. The
cache-blocking analysis determines $T_{blk}$, $C_{blk}$, and $K_{blk}$
under the L1 and L2 cache-capacity constraints to limit data movement.

\subsubsection{Micro-kernel design}
With the packed layouts shown in Fig.~\ref{fig:fig_3}, each GEMM
micro-kernel uses $\alpha$ and $\eta/4$ vector registers for the
transformed input and filter operands, respectively, together with
$\alpha\eta/4$ accumulator registers. To overlap operand loading with
computation, the ping-pong schedule~\cite{ref22, ref23, ref24, ref25}
reserves an additional $\alpha+\eta/4$ registers for preloading the
next operands. The total number of registers must therefore satisfy:
\begin{equation}
  2\alpha + \eta/2 + \alpha \times \eta/4 \leq 32.
  \label{eq:8}
 \end{equation}  

 We set the parameter $\eta$ to satisfy the constraint specified in Equation~\eqref{eq:9}, 
 enabling efficient SIMD loading for FP32 precision. 
 Considering that the filter dimension $K$ in convolutional neural networks is typically a multiple of 16, 
 we accordingly configure the blocking size along the $K$ dimension to also be a multiple of 16. 
 This deliberate design choice, as detailed in Subsection~\ref{subsec:block_size}, 
 effectively mitigates potential edge cases along the $K$ dimension that might otherwise degrade the arithmetic intensity of the GEMM operation.
 \begin{equation}
  \eta \% 4 = 0.
  \label{eq:9}
 \end{equation}  

 Over $\theta$ reduction iterations, the micro-kernel executes $\alpha$
 input loads, $\eta$ filter loads ($\eta/\theta$ per iteration), and
 $\alpha\eta$ scalar-vector FMA instructions. Counting each FMA as two
 floating-point operations, the resulting average CMR is:
 \begin{equation}
  \frac{2 \times \alpha \times \eta}{\alpha + \eta}.
  \label{eq:10}
 \end{equation}

 Maximizing Equation~\eqref{eq:10} subject to
 Equations~\eqref{eq:8} and~\eqref{eq:9} yields
 $(\alpha,\eta)=(7,8)$. As shown in Table~\ref{tab:table1}, $T$
 generally decreases while $C$ and $K$ increase across the evaluated
 layers, which raises the relative cost of edge cases along the $T$
 dimension. We therefore also implement a $(4,16)$ micro-kernel, whose
 smaller $\alpha$ reduces the $T$-dimension remainder at the cost of a
 lower CMR. The implementation selects $(4,16)$ when both $C$ and $K$
 exceed $T$. Otherwise, it uses $(7,8)$.

\begin{figure}[t]
  \centering
  \subfloat[SIMD register allocation for the $(4,16)$ micro-kernel]
  {\includegraphics[width=0.8\linewidth]{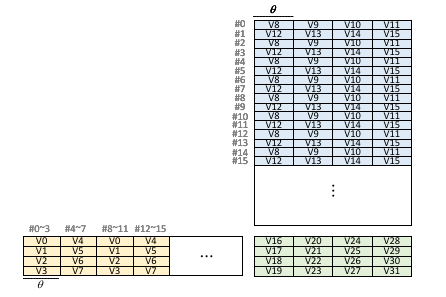}
  \label{fig:fig_4a}}\\
  \subfloat[SIMD register allocation for the $(7,8)$ micro-kernel]
  {\includegraphics[width=0.7\linewidth]{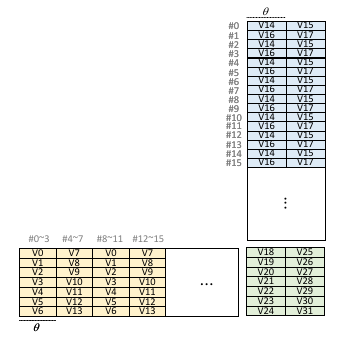}
  \label{fig:fig_4b}}
  \caption{Physical register allocation and stage-specific operand
  refilling in the GEMM micro-kernels. The notation $\#s$ denotes
  reduction stage $s$. Yellow, blue, and green represent transformed
  inputs, transformed filters, and accumulators, respectively.
  Both configurations occupy all 32 SIMD registers.}
  \label{fig:fig_4}
\end{figure}

Figure~\ref{fig:fig_4} maps the register constraint in
Equation~\eqref{eq:8} to the physical SIMD registers and shows how the
registers holding the transformed input and transformed filter operands
alternate across reduction stages. In the $(4,16)$ micro-kernel shown
in Fig.~\ref{fig:fig_4a}, \texttt{v0} to \texttt{v7} hold the
transformed input operands, \texttt{v8} to \texttt{v15} hold the
transformed filter operands, and \texttt{v16} to \texttt{v31} hold the
16 accumulators. In the $(7,8)$ micro-kernel shown in
Fig.~\ref{fig:fig_4b}, the corresponding register ranges are
\texttt{v0} to \texttt{v13}, \texttt{v14} to \texttt{v17}, and
\texttt{v18} to \texttt{v31}, with 14 accumulator registers. The
entries indexed by $\#s$ show which input and filter registers are
used at each reduction stage and how their roles alternate.

Algorithm~\ref{alg:gemm_microkernel} presents the common schedule
independently of these physical register numbers. At each stage,
$V_{\mathrm{cur}}$ and $U_{\mathrm{cur}}$ denote the registers holding
the transformed input and transformed filter operands, respectively, for the current AArch64
\texttt{FMLA} (floating-point fused multiply-add to accumulator)
instructions, while $V_{\mathrm{nxt}}$ and $U_{\mathrm{nxt}}$ denote
the registers being filled for subsequent stages.

For each $i$, $V_{\mathrm{cur}}[i]$ is a vector register
containing four packed transformed input values for the current group
of reduction coordinates. $V_{\mathrm{cur}}[i,\ell]$ denotes the
value in lane $\ell$, where $\ell\in\{0,1,2,3\}$. During lane stage
$\ell$, $U_{\mathrm{cur}}[j]$ contains four packed
transformed filter values for local vector column $j$ at the
corresponding reduction coordinate.

The vector-by-element form of the AArch64 \texttt{FMLA} instruction
multiplies the four values in $U_{\mathrm{cur}}[j]$ by the scalar
$V_{\mathrm{cur}}[i,\ell]$ and accumulates the products into
$\hat{O}[i,4j{:}4j+3]$. The lane selection is encoded by the
instruction, so no separate broadcast is required. At each lane
stage, the micro-kernel updates $\alpha\eta/4$ distinct accumulator
vectors. The $(4,16)$ kernel therefore maintains 16 independent FMLA
chains, whereas the $(7,8)$ kernel maintains 14. The schedule visits
all independent accumulators before returning to the same accumulator
at the next reduction coordinate, increasing the instruction distance
between dependent updates.

\begin{algorithm}[t]
  \small
  \caption{Schematic steady-state schedule of the NEON GEMM
  micro-kernels}
  \label{alg:gemm_microkernel}
  \SetAlgoNoEnd
  \SetAlgoLined

  Initialize or reload the accumulator vectors of $\hat{O}$ using
  \texttt{eor} or \texttt{ld1}\;
  Load $V_{\mathrm{cur}}$ and $U_{\mathrm{cur}}$ using \texttt{ldr}\;

  \For{$c=0$ \KwTo $C_{blk}-1$ \KwSty{step} $4$}{
    \For{$\ell=0$ \KwTo $3$}{
      $\hat{O}[i,4j{:}4j+3]\leftarrow
      \operatorname{FMLA}\bigl(
      \hat{O}[i,4j{:}4j+3],
      U_{\mathrm{cur}}[j],
      V_{\mathrm{cur}}[i,\ell]
      \bigr)$,
      $\ \forall\,0\leq i<\alpha,\;
      0\leq j<\eta/4$\;

      Fill $U_{\mathrm{nxt}}$ and the scheduled part of
      $V_{\mathrm{nxt}}$ using \texttt{ldr}\;
      $\operatorname{swap}
      (U_{\mathrm{cur}},U_{\mathrm{nxt}})$\;
    }
    $\operatorname{swap}
    (V_{\mathrm{cur}},V_{\mathrm{nxt}})$\;
  }

  Store the accumulator vectors using \texttt{st1}\;
\end{algorithm}

Algorithm~\ref{alg:gemm_microkernel} shows the steady-state
instruction sequence. The prologue, epilogue, and tail cases are
omitted because they do not change the register allocation or the
operand-register switching schedule.

The stage-specific register-refilling pattern shown in
Fig.~\ref{fig:fig_4} uses post-indexed \texttt{ldr} instructions. In
the $(4,16)$ kernel, each of the four lane stages
loads one register of $V_{\mathrm{nxt}}$ and four registers of
$U_{\mathrm{nxt}}$. These loads are distributed among the 16
independent FMLA instructions. In the $(7,8)$ kernel, the seven
registers of $V_{\mathrm{nxt}}$ are filled according to a
$2,2,2,1$ schedule over the four lane stages. Each stage also loads
two registers of $U_{\mathrm{nxt}}$ before issuing its 14 FMLA
instructions. The steady-state loop unrolls two consecutive
four-lane reduction groups.

 \subsubsection{Cache blocking analysis}
 \label{subsec:block_size}
 Our approach employs a heuristic-based method to determine the blocking size during the instantiation phase, 
 with a design principle grounded in the cache capacity to minimize data movement overhead. 
 This strategy ensures that the blocks are appropriately sized to optimize performance. 
 During the iteration in the $C$ direction, as outlined in line 13 of Algorithm \ref{alg:alg1}, the $T_{blk} \times K_{blk}$ block of output should be consistently retained in the L2 cache. 
 Additionally, the L2 cache must accommodate the $T_{blk} \times C_{blk}$ block of input and the $C_{blk} \times K_{blk}$ block of filter for the current iteration. 
 Moreover, it should prefetch the necessary blocks for the subsequent iteration. 
 
 We denote the cache capacity as $\mathbb{C}$. Thus, the blocking parameters must be carefully chosen to satisfy the following constraint:
\begin{equation}
  T_{blk} \times K_{blk} + 2 \times (T_{blk} \times C_{blk} + C_{blk} \times K_{blk}) < \mathbb{C}_{L2}.
  \label{eq:11}
\end{equation}

In GEMM kernel, our design prioritizes the processing of $T_{blk}/\alpha$ blocks of size $\alpha \times C_{blk}$ for the input, 
utilizing the same  $C_{blk} \times \eta$ block of the filter for the micro-kernel.  
Furthermore, the corresponding $T_{blk} \times K_{blk}$ block of results is retained in the L1 cache. 

To ensure that the L1 cache can accommodate the necessary data blocks, the capacity constraint is given by
\begin{equation}
  T_{blk} \times K_{blk} + 2 \times\alpha \times C_{blk} + C_{blk} \times \eta < \mathbb{C}_{L1}.
  \label{eq:12}
\end{equation}

We use $\mathbb{B}$ to denote the data transfer bandwidth of each level of the memory hierarchy, where $\mathbb{B}_{M}$ represents the data transfer bandwidth of the levels below the L2 cache (i.e., the LLC and main memory).
Thus, the total data movement overhead for input, filter, and output can be modeled as follows:
\newcommand{\cDot}{\mathbin{\mkern-3mu\cdot\mkern-3mu}}

\begin{subequations}
  \allowdisplaybreaks    %
  \begin{IEEEeqnarray}{rCl}
    \mathbb{O}_{input} 
    &\approx& 
    \tfrac{T}{T_{blk}} \cDot \tfrac{K}{K_{blk}}\cDot \mathsmaller{\mathsmaller{L}} \cDot \tfrac{C}{C_{blk}} \cDot (\tfrac{T_{blk}}{\alpha}\cDot \tfrac{K_{blk}}{\eta} \cDot \mathsmaller{\mathsmaller{\alpha}} \cDot \mathsmaller{\mathsmaller{C_{blk}}})\cDot(\tfrac{1}{\mathbb{B}_{L1}} \mathsmaller{\mathsmaller{+}} \tfrac{1}{\mathbb{B}_{L2} })
    \nonumber\\
    &&\hphantom{\approx}\; \mathsmaller{\mathsmaller{+}} \tfrac{T}{T_{blk}} \cDot \tfrac{K}{K_{blk}}\cDot \mathsmaller{\mathsmaller{L}} \cDot \tfrac{C}{C_{blk}} \cDot \tfrac{T_{blk}\cdot C_{blk}}{\mathbb{B}_{M}}
   \nonumber\\
    &=&
      \tfrac{T\cdot K \cdot L \cdot C}{\eta} \cDot(\tfrac{1}{\mathbb{B}_{L1}} \mathsmaller{\mathsmaller{+}} \tfrac{1}{\mathbb{B}_{L2} }) \mathsmaller{\mathsmaller{+}} \tfrac{T\cdot K \cdot L \cdot C}{K_{blk}\cdot \mathbb{B}_{M}}, 
      \IEEEeqnarraynumspace               
    \label{eq:13a}
  \end{IEEEeqnarray}
  \begin{IEEEeqnarray}{rCl}
  \mathbb{O}_{filter} &\approx&  \tfrac{T}{T_{blk}} \cDot \tfrac{K}{K_{blk}}\cDot \mathsmaller{\mathsmaller{L}} \cDot \tfrac{C}{C_{blk}} \cDot (\tfrac{K_{blk}}{\eta} \cDot \mathsmaller{\mathsmaller{C_{blk}}} \cDot\mathsmaller{\mathsmaller{\eta}})\cDot(\tfrac{1}{\mathbb{B}_{L1}} \mathsmaller{\mathsmaller{+}} \tfrac{1}{\mathbb{B}_{L2} })
  \nonumber\\
  && \hphantom{\approx}\,\mathsmaller{\mathsmaller{+}} \tfrac{T}{T_{blk}} \cDot \tfrac{K}{K_{blk}}\cDot \mathsmaller{\mathsmaller{L}} \cDot \tfrac{C}{C_{blk}} \cDot \tfrac{C_{blk}\cdot K_{blk}}{\mathbb{B}_{M}}
  \nonumber\\
                  &=& \tfrac{T\cdot K \cdot L \cdot C}{T_{blk}} \cDot(\tfrac{1}{\mathbb{B}_{L1}} \mathsmaller{\mathsmaller{+}} \tfrac{1}{\mathbb{B}_{L2} } \mathsmaller{\mathsmaller{+}} \tfrac{1}{\mathbb{B}_{M}}),
                  \IEEEeqnarraynumspace
\label{eq:13b}
  \end{IEEEeqnarray}
  \begin{IEEEeqnarray}{rCl}
    \mathbb{O}_{output} &\approx&  \tfrac{T}{T_{blk}} \cDot \tfrac{K}{K_{blk}}\cDot \mathsmaller{\mathsmaller{L}}  \cDot T_{blk} \cDot K_{blk} \cDot (\tfrac{1}{\mathbb{B}_{L2}} \mathsmaller{\mathsmaller{+}} \tfrac{1}{\mathbb{B}_{M} })
    \nonumber\\
&& \hphantom{\approx}\,+ \tfrac{T}{T_{blk}} \cDot \tfrac{K}{K_{blk}}\cDot \mathsmaller{\mathsmaller{L}}  \cDot \tfrac{C}{C_{blk}} \cDot \tfrac{T_{blk}}{\alpha} \cDot \tfrac{K_{blk}}{\eta} \tfrac{\alpha\cdot\eta}{\mathbb{B}_{L1}}
\nonumber\\
                  &=& \mathsmaller{\mathsmaller{T}}\cDot \mathsmaller{\mathsmaller{K}} \cDot \mathsmaller{\mathsmaller{L}} \cDot [(\tfrac{1}{\mathbb{B}_{L2}} + \tfrac{1}{\mathbb{B}_{M} }) + \tfrac{C}{C_{blk} \cdot \mathbb{B}_{L1}}].
                  \IEEEeqnarraynumspace  
                  \label{eq:13c}
  \end{IEEEeqnarray}
\end{subequations}

Therefore, we can calculate the total data movement overhead $\mathbb{O}$ by summarizing Equations \eqref{eq:13a}, \eqref{eq:13b} and \eqref{eq:13c} into the following expression:

\begin{IEEEeqnarray}{rCl}
  \mathbb{O} & = & \mathbb{O}_{\text{input}}
                 + \mathbb{O}_{\text{filter}}
                 + \mathbb{O}_{\text{output}} \nonumber\\
             & = & T\cDot K\cDot L \cDot C \cDot \Biggl[
                   \frac{1}{\mathbb{B}_{L1}}\cDot\left(
                     \frac{1}{\eta} + \frac{1}{T_{blk}} + \frac{1}{C_{blk}}
                   \right) \nonumber\\
             &&    +\,\frac{1}{\mathbb{B}_{L2}}\cDot\left(
                     \frac{1}{\eta} + \frac{1}{T_{blk}} + \frac{1}{C}
                   \right) \nonumber\\
             &&    +\,\frac{1}{\mathbb{B}_{M}}\cDot\left(
                     \frac{1}{\eta K_{blk}} + \frac{1}{T_{blk}}
                     + \frac{1}{C}
                   \right)
                 \Biggr].
  \label{eq:14}
  \end{IEEEeqnarray}

The strategy of our method focuses on minimizing Equation \eqref{eq:14}  under the constraints specified in Equations \eqref{eq:11} and \eqref{eq:12}. 
Considering the characteristics of the channels in convolutional neural networks, both $K_{blk}$ and $C_{blk}$ are configured to be divisible by 16. 
This configuration ensures that  $\eta$ satisfies Equation \eqref{eq:9}, thereby circumventing potential edge cases that could arise in the  $K$ and $C$ dimension. 
By setting $K_{blk}$ and $C_{blk}$ to be multiples of 16, we can streamline the computational process, enhancing both the efficiency and stability of the GEMM operation.

\subsection{Output Transformation}
After executing computations in the Winograd domain, the results need to be transformed back into the spatial domain. 
The number of inverse transformations is proportional to $T\times K$,
which decreases across the deeper layers considered in this work.
This procedure in our approach is analogous to the input and filter transformations demonstrated in Subsection \ref{subsec: input_filter_transformation}, where each SIMD register holds $\theta$ elements simultaneously.
To exploit cache locality, the output transformation kernel is invoked as soon as the computation of the results $L \times T_{blk}  \times K_{blk}$  is complete, 
when these results are stored in a temporary array, as outlined in Algorithm \ref{alg:alg1}.
The data layout of this temporary array is structured as  [$L$][$K_{blk}/\eta$][$T_{blk}/\alpha$][$\eta/\theta$][$\alpha$][$\theta$], which ensures continuous memory access for GEMM storage and facilitates efficient loading during the inverse transformation. 
Subsequently, the results in the spatial domain are stored back into the main memory following the standard convolutional neural network format [$N$][$K$][$P$][$Q$].

\subsection{Parallel Strategies}
We employ OpenMP \cite{ref26} to parallelize the fused framework presented in Algorithm~\ref{alg:alg1}, 
utilizing static scheduling to pre-allocate workloads to each available core/thread. 

The primary objective is to achieve a balanced distribution of the computational workload across threads for convolutional layers of varying sizes, 
thereby reducing core idling, minimizing cache contention and synchronization overhead, enhancing both parallel efficiency and scalability.

To fully exploit the ARM architecture and achieve these goals, we adopt a multi-dimensional parallelization strategy  which combines inter-batch and intra-batch parallelization. 
Specifically, we parallelize the loops corresponding to dimensions in lines 7, 8, 9, and 11 of Algorithm 1, as well as the outermost batch dimension. 
The dimensions involved in parallelization include $N$, $T$ and $C/K$, which together form a three-level parallelism structure. Correspondingly, 
the available $P$ threads are divided into a three-dimensional grid of $P_{N} \times P_{T} \times P_{C/K}$. 
By collapsing  lines 8 and 9 of Algorithm \ref{alg:alg1} along the $C$ dimension, 
the total number of sub-tasks is the product of the sub-task counts in the $N$, $T$, and $C$/$K$ dimensions, given by $N \times \frac{T}{T_{blk}} \times \left( \frac{C}{C_T} + \frac{K}{K_{blk}} \right)$.
To ensure that each thread is sufficiently utilized and that the total number of subtasks is evenly divisible across threads,
the maximum thread counts for the $N$, $T$, and $C$/$K$ dimensions are empirically set to $max(1,\frac{N}{2})$, $max(1, 2^{\left\lfloor \log_2\left(\frac{T/T_{blk}}{4}\right) \right\rfloor})$, and $max(\frac{min(C/C_T, K/K_{blk})}{4},1)$, respectively. 
For convenience, these values will be referred to as $P^{max}_N$, $P^{max}_T$, and $P^{max}_{C/K}$ in subsequent discussions.

Given the aim of maximizing the independence of sub-tasks executed within the same thread/core, 
the parallelism priority is set as $\mathcal{P}(N)>\mathcal{P}(T)>\mathcal{P}(C/K)$, 
where $\mathcal{P}(X)$ denotes the priority of dimension $X$.
Therefore, the thread allocation across dimensions in the thread grid can be recursively determined as $P_N=min(P,P^{max}_N)$, $P_T=min(\frac{P}{P_N}, P^{max}_T)$, and $P_{C/K}=\frac{P}{P_N \times P_T}$. 
When $P \leq P^{max}_N$, only inter-batch parallelism is performed; otherwise, intra-batch parallelism is enabled simultaneously. 
Intra-batch parallelism can be further categorized into three modes based on problem size: the $T$ mode for layers with a sufficient number of tiles to offer adequate parallelism, the multi-dimensional mode when the thread count exceeds the maximum number of threads in the T-dimension, 
and the $C$/$K$ mode for layers where T is relatively small and not worth parallelizing, with $T_{blk}$ set to $T$. 

We implement static workload distribution within a single parallel region to mitigate overhead due to thread allocation and destruction in the inner loops introduced by OpenMP nested parallelism. 
The $\frac{C}{C_T}$ sub-tasks of the input transformation and the $\frac{K}{K_{blk}}$ sub-tasks of the fused GEMM and output transformation are executed in parallel by unit thread groups, 
each consisting of $P_{C/K}$ threads. $P_T$ unit thread groups form a composite group $(P_{T} \times P_{C/K}\text{ threads})$ to perform intra-batch parallelization. 
Likewise, $P_{N}$ composite groups enable parallelism along batches. Synchronization within the unit thread group is required after input transformation. 
We employ an intra-group synchronization mechanism based on sense variables and OpenMP atomic operations to enhance the efficiency while ensuring its correctness.

\subsection{Extensions to Other Precisions and ISAs}
The block-wise execution schedule and thread decomposition are
independent of arithmetic precision. Prior work has demonstrated
FP16 Winograd convolution on ARM many-core processors~\cite{ref46}.
An FP16 extension could
retain the FMLA-based kernel structure by increasing $\theta$ and
retuning the packed layout and blocking parameters. INT8 requires
separate micro-kernels because the current FMLA schedule does not
directly map to base-NEON integer operations. A high-throughput
implementation would use dot-product or I8MM extensions, depending
on processor support~\cite{ref47}, with additional treatment to preserve
numerical accuracy in the Winograd transforms~\cite{ref33,ref34}.

On ARM cores supporting the AArch64/NEON instructions used by our
implementation, the existing assembly micro-kernels can be reused,
with performance tuning focused on instruction scheduling, prefetch
distances, cache blocking, and thread allocation. Using SVE instead
of NEON requires new transformation and GEMM micro-kernels, together
with register-blocking and packed-layout adjustments for its
vector-length and predication mechanisms. Porting to non-ARM
architectures requires reimplementing these micro-kernels using the
target SIMD instructions and adapting the associated layouts and
blocking parameters. The block-wise execution structure and
thread-decomposition strategy can be retained in each case.

\section{EXPERIMENTS}
\label{sec:experiments}

\subsection{Experimental Setup}
\label{subsec:experimental_setup}

We evaluate the performance and accuracy of our method against
NCNN~\cite{ref12} (commit \texttt{fafb897ff7f6}),
NNPACK~\cite{ref13} (commit \texttt{70a77f485e8b}),
FastConv~\cite{ref14} (commit \texttt{dbae5da3c1710}),
and ACL~\cite{ref15} (v24.11, commit \texttt{f44f09dfd263}).
 As summarized in Table~\ref{tab:table1}, we benchmark the widely adopted VGG-16 \cite{ref2} and FusionNet \cite{ref3}, 
 which together represent a range of network scales.
 The experiments were conducted on three distinct ARMv8-based platforms. Specifically, 
 we employed (1) a Kunpeng 920 processor \cite{ref27} with 64 KB L1 instruction/data caches, 
 a 512 KB L2 cache, and a 64 MB shared L3 cache; (2) an AWS Graviton2 M6g instance \cite{ref28} with 64 KB L1 instruction/data caches,
 a 1 MB L2 cache, and a 32 MB shared L3 cache; and (3) a Phytium 2000+ processor \cite{ref29} with 64 KB L1 instruction/data caches and a 2 MB L2 cache shared by a four-core cluster,
 but no L3 cache.

 NCNN dynamically selects among three Winograd convolution variants—$F(2\times2,3\times3)$, $F(4\times4,3\times3)$, and $F(6\times6,3\times3)$—based on the problem scale, 
 and implements these configurations using GEMM. 
 NNPACK adopts only the $F(6\times6,3\times3)$ configuration with Tuple Element-Wise Multiplication (TEWMM),
 which exhibits lower arithmetic intensity than GEMM while significantly reducing transformation overhead.
 FastConv, a high-performance library optimized for ARM CPUs, implements $F(6\times6,3\times3)$ Winograd convolution through TensorGEMM,
 a GEMM-like algorithm. Although TensorGEMM shares a similar arithmetic intensity expression with GEMM, 
 under the same vector register constraints it demonstrates lower arithmetic intensity, and similarly mitigates transformation costs.
 Finally, ACL employs an $F(4\times4,3\times3)$ Winograd convolution with GEMM and performs additional data packing after transformation. We evaluate ACL using the NCHW layout.

 For each comparison, all implementations use the same thread count and are restricted to the same set of CPU cores.
 To minimize stochastic fluctuations, each evaluation is preceded by two warm-up executions to stabilize the cache state, 
 followed by five measured runs whose average is reported as the final result.

  \begin{table}[h!]
    \centering
    \caption{Different convolution neural networks.}
    \label{tab:table1}
    \setlength{\tabcolsep}{4pt}
    \begin{tabular}{|c|c|c|c|c|}
      \hline
      \textbf{Layer} & \textbf{C} & \textbf{K} & \textbf{H \& W} & \textbf{R \& S} \\ \hline
      VggNet\_1.2   &  64  &  64  & 224 & 3 \\ \hline
      VggNet\_2.2   & 128  & 128  & 112 & 3 \\ \hline
      VggNet\_3.2   & 256  & 256  &  56 & 3 \\ \hline
      VggNet\_4.2   & 512  & 512  &  28 & 3 \\ \hline
      VggNet\_5.2   & 512  & 512  &  14 & 3 \\ \hline
      FusionNet\_1.2&  64  &  64  & 640 & 3 \\ \hline
      FusionNet\_2.2& 128  & 128  & 320 & 3 \\ \hline
      FusionNet\_3.2& 256  & 256  & 160 & 3 \\ \hline
      FusionNet\_4.2& 512  & 512  &  80 & 3 \\ \hline
      FusionNet\_5.2&1024  &1024  &  40 & 3 \\ \hline
    \end{tabular}
  \end{table}

  \subsection{Ablation Test}
  \label{subsec:ablation}

  \begin{figure}[t]
    \centering
    \subfloat[Comparison of input transformation data reuse efficiency for $F(2^2,3^2)$]{\includegraphics[width=1\linewidth]{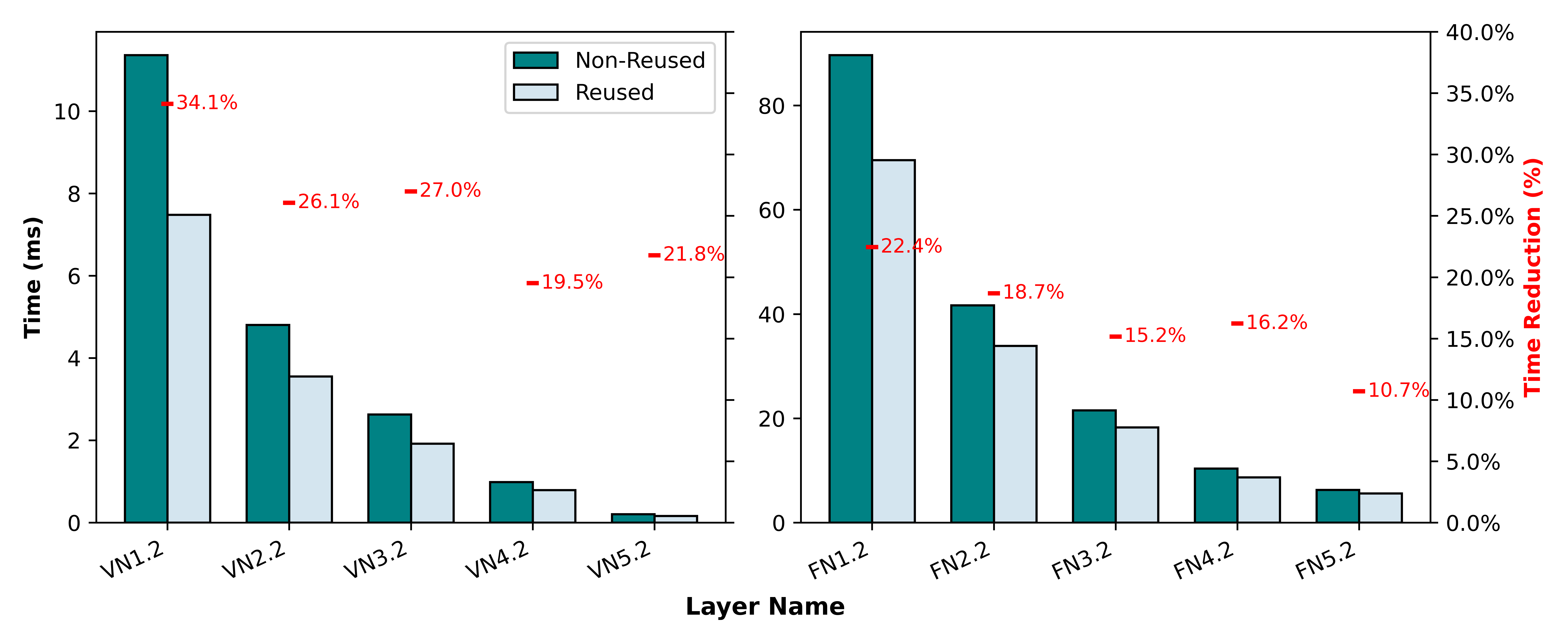}\label{fig:fig_5a}} \\
    \subfloat[Comparison of data prefetching efficiency during transformation]{\includegraphics[width=1\linewidth]{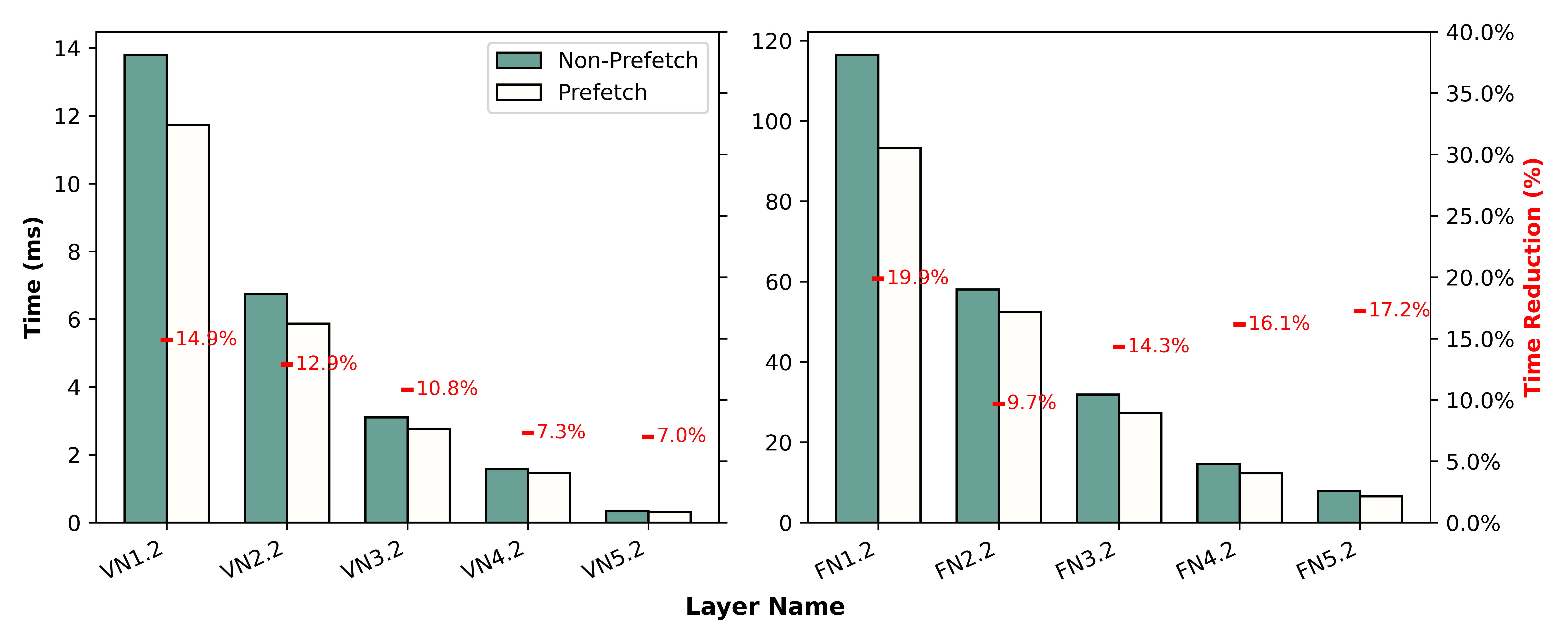}\label{fig:fig_5b}}\\
    \subfloat[Comparison of LLC miss rates between the fused block-wise and stage-wise non-fused schedules]{\includegraphics[width=1\linewidth]{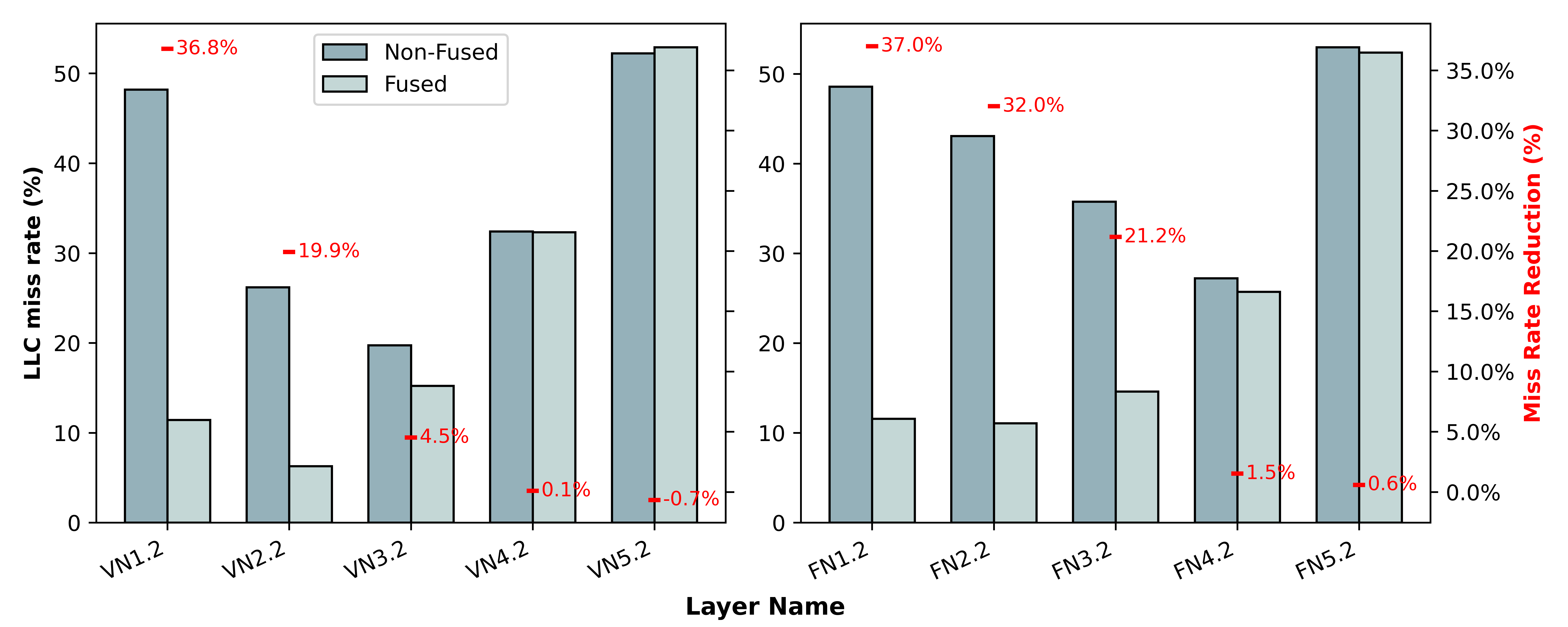}\label{fig:fig_5c}}
    \caption{Layer-wise performance evaluation demonstrating the effectiveness of the proposed optimization techniques: 
    (a) data reuse efficiency for $F(2^2,3^2)$, 
    (b) software prefetching efficiency, 
    and (c) LLC miss rate comparison between the fused block-wise and stage-wise non-fused schedules.}
    \label{fig:fig_5}
  \end{figure}

  \begin{figure}[t]
    \centering
    \subfloat[Assembly versus intrinsics for transformation]{\includegraphics[width=1\linewidth]{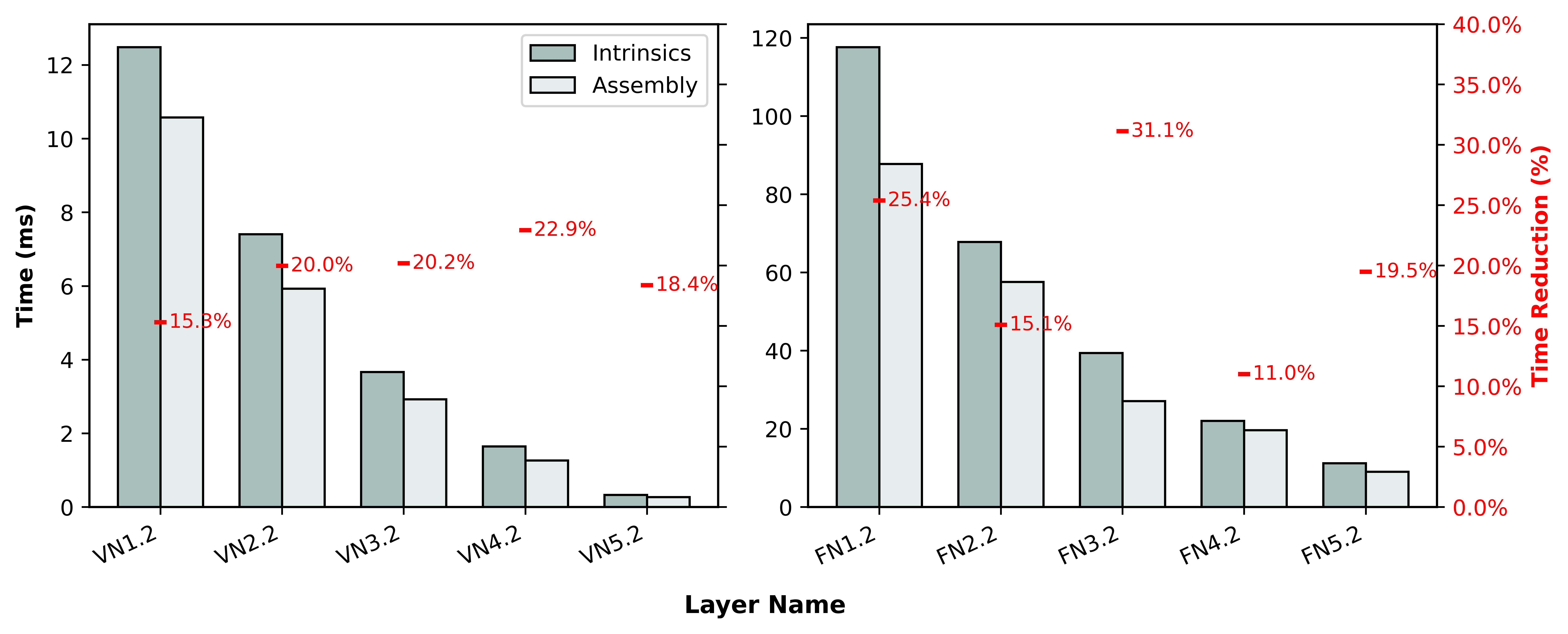}\label{fig:fig_6a}}\\
    \subfloat[Assembly versus intrinsics for GEMM]{\includegraphics[width=1\linewidth]{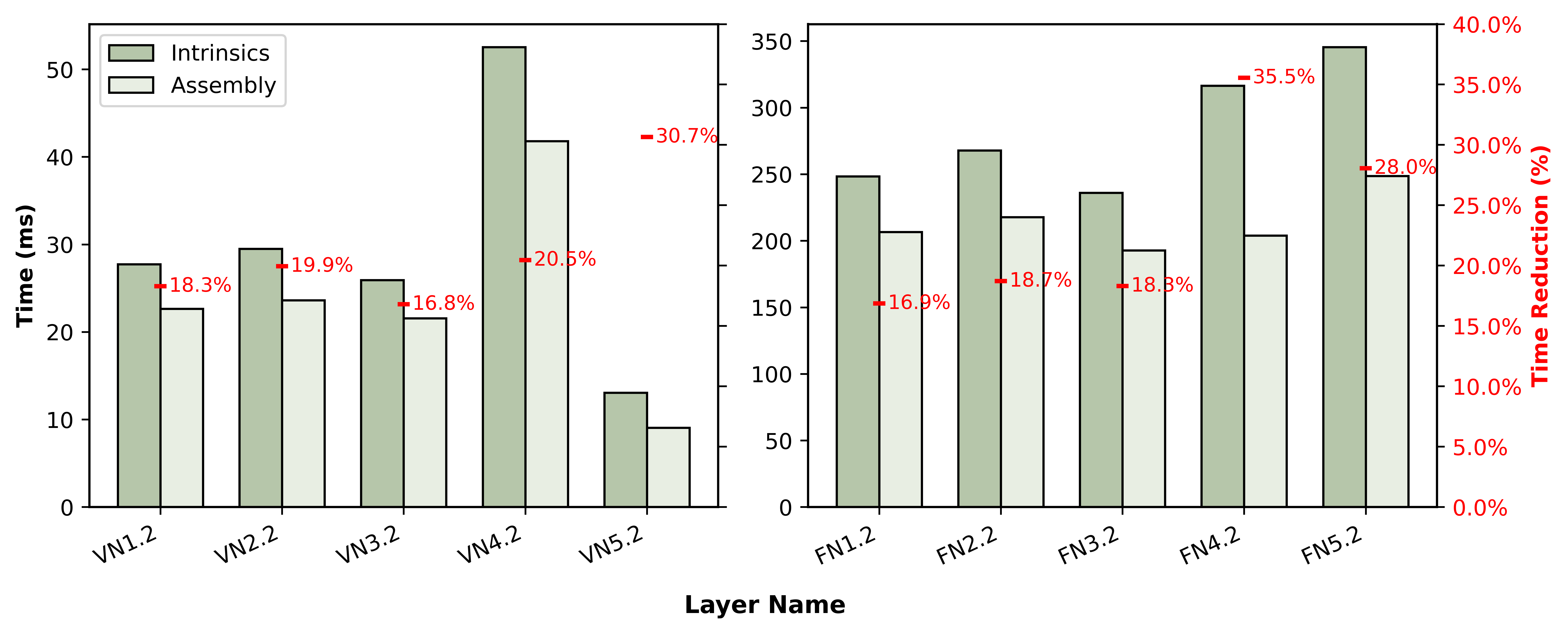}\label{fig:fig_6b}}
    \caption{Layer-wise single-thread runtime comparison between the hand-written AArch64 assembly and semantically equivalent ARM NEON intrinsics implementations: (a) the transformation stage, combining input and output transformations, and (b) the GEMM stage. The red annotations show the runtime reduction achieved by assembly relative to intrinsics.}
    \label{fig:fig_6}
  \end{figure}

To comprehensively evaluate the optimization effectiveness of our proposed techniques, 
we performed detailed ablation studies on the Kunpeng 920 platform. 
All evaluations were conducted layer-wise across the benchmarked convolutional layers, 
ensuring identical execution parameters between the optimized and baseline versions,
with optimization techniques being the sole variable.

\noindent\textbf{\textit{Input transformation data reuse.}}
We evaluated the runtime performance of the input transformation phase for the $F(2\times2,3\times3)$ configuration, 
comparing implementations with and without data reuse, as illustrated in Fig.~\ref{fig:fig_5a}. 
The implementation employing data reuse achieved notable runtime reductions—approximately $34.1\%$ for VN1.2 and $22.4\%$ for FN1.2—relative to the baseline implementation. 
These two layers represent the most computationally intensive input transformation phases in their respective networks. 
Additionally, the observed runtime reductions exhibit a decreasing trend as the spatial dimension ($H$) diminishes, 
primarily due to the reduced availability of reusable data elements.

\noindent\textbf{\textit{Software prefetching for transformations.}}
We evaluated the efficiency gains obtained from the manual insertion of software prefetching instructions designed to mitigate performance penalties associated with strided memory accesses. 
Considering that the GEMM micro-kernel performs consecutive memory accesses, 
our analysis focused exclusively on the input and output transformation stages. 
Fig.~\ref{fig:fig_5b} illustrates that software prefetching consistently reduced transformation runtimes by $7.0\%$ to $19.9\%$ compared to the non-prefetch implementation.

\noindent\textbf{\textit{Fused block-wise execution schedule.}}
We evaluated LLC miss rates to assess the fused block-wise schedule relative to the stage-wise non-fused schedule, 
which separately executes the input transformation, GEMM computation, and output transformation stages. Results shown in Fig.~\ref{fig:fig_5c} demonstrate a substantial reduction in LLC miss rates across most layers when utilizing the fused block-wise schedule, 
indicating superior cache locality. Additionally, the fused block-wise schedule maintained competitive performance advantages even in deeper layers of the networks, 
reinforcing its effectiveness for a wide range of scenarios.

\begin{table*}[t]
\centering
\caption{Controlled scheduling ablation on the Kunpeng 920 ($N=1$, 16 threads).}
\label{tab:scheduling_ablation}
\begingroup
\footnotesize
\setlength{\tabcolsep}{5pt}
\renewcommand{\arraystretch}{1.15}
\begin{tabular*}{\textwidth}{@{\extracolsep{\fill}}lrrrrrrrrr@{}}
\toprule
& \multicolumn{3}{c}{\shortstack{Latency\\(ms)}} & \multicolumn{3}{c}{\shortstack{Estimated intermediate\\storage (MiB)}} & \multicolumn{3}{c}{\shortstack{Barrier waiting\\(\%)}} \\
\cmidrule(lr){2-4}\cmidrule(lr){5-7}\cmidrule(lr){8-10}
Layer & A & B & C & A & B & C & A & B & C \\
\midrule
VN1.2 & 10.582 & 4.886 & \textbf{4.692} & 42.78 & 6.56 & \textbf{2.62} & 6.67/\textbf{1.05} & \textbf{0.00}/12.74 & 2.86/2.65 \\
VN2.2 & 6.230 & 6.008 & \textbf{3.645} & 22.56 & 12.55 & \textbf{4.38} & 34.81/9.23 & \textbf{0.00}/43.20 & 2.58/\textbf{0.96} \\
VN3.2 & 15.760 & 15.783 & \textbf{3.167} & 10.12 & \textbf{5.41} & 9.62 & 86.11/\textbf{4.36} & \textbf{0.00}/90.49 & 2.04/20.09 \\
VN4.2 & 12.921 & 12.424 & \textbf{3.492} & 10.56 & 5.91 & \textbf{5.25} & 74.55/\textbf{2.37} & \textbf{0.00}/76.96 & 3.03/3.82 \\
VN5.2 & 9.401 & 9.412 & \textbf{0.737} & 2.25 & \textbf{1.20} & 2.25 & 92.79/0.89 & \textbf{0.00}/93.70 & 4.24/\textbf{0.66} \\
FN1.2 & 92.238 & 28.959 & \textbf{28.786} & 357.78 & \textbf{5.91} & \textbf{5.91} & 1.44/\textbf{0.33} & \textbf{0.00}/2.35 & \textbf{0.00}/2.11 \\
FN2.2 & 50.954 & 26.220 & \textbf{25.965} & 175.56 & \textbf{10.00} & \textbf{10.00} & 1.18/\textbf{0.38} & \textbf{0.00}/1.61 & \textbf{0.00}/1.64 \\
FN3.2 & 41.744 & 35.298 & \textbf{22.982} & 91.12 & 31.50 & \textbf{7.00} & 17.12/2.65 & \textbf{0.00}/19.15 & 2.72/\textbf{0.62} \\
FN4.2 & 115.413 & 114.072 & \textbf{23.923} & 56.25 & \textbf{31.64} & 56.25 & 83.98/3.53 & \textbf{0.00}/87.50 & 0.64/\textbf{0.02} \\
FN5.2 & 264.958 & 264.548 & \textbf{24.857} & 28.12 & \textbf{14.94} & 28.12 & 92.85/0.89 & \textbf{0.00}/93.75 & 0.71/\textbf{0.02} \\
\bottomrule
\end{tabular*}
\endgroup
\par\smallskip
\begin{minipage}{\textwidth}\footnotesize
A: stage-wise execution with thread assignment along the $T$ dimension; B: block-wise execution with the same thread assignment; C: block-wise execution with thread decomposition across $T$ and $C/K$. All variants use identical micro-kernels, layouts, and blocking parameters. Each waiting entry gives internal/final barrier waiting. Each waiting percentage is computed as $100\sum_{p=1}^{P}t_{p,\mathrm{wait}}/(P\,t_{\mathrm{profile}})$, where $P$ is the thread count, $t_{p,\mathrm{wait}}$ is the accumulated waiting time of thread $p$ for the corresponding barrier type, and $t_{\mathrm{profile}}$ is the wall-clock time of the same instrumented run. We report the median percentage over three runs. MiB=$2^{20}$ bytes. Bold values mark per-layer minima (including ties); internal and final waiting are compared separately.
\end{minipage}
\end{table*}

\noindent\textbf{\textit{Execution scheduling and thread decomposition.}}

We compare three execution variants: (A) stage-wise execution with
thread assignment along the $T$ dimension, (B) block-wise execution with
the same thread assignment, and (C) block-wise execution with thread decomposition
across $T$ and $C/K$. All variants use identical AArch64 assembly
micro-kernels, packed layouts, Winograd variants, and blocking parameters.
Table~\ref{tab:scheduling_ablation} reports the results for the ten layers
at $N=1$ using 16 threads on the Kunpeng 920.

We estimate the intermediate storage required between the input
transformation, GEMM, and output transformation stages. For A, we count
the full-layer intermediate storage; for B and C, we estimate an upper
bound on the storage for concurrently active blocks.
Barrier waiting is reported as internal/final waiting: the first
value measures waiting at barriers during computation, and the second
measures waiting at the final barrier after each thread completes its
assigned work. Both are expressed as percentages of aggregate thread time
and include barrier execution and unequal arrival times.

Across the ten layers, B achieves a geometric mean speedup of
1.33$\times$ over A. C achieves geometric mean speedups of
2.81$\times$ over B and 3.74$\times$ over A.

Block-wise execution reduces intermediate storage by reusing buffers
across tile blocks, so only the concurrently active blocks need to
retain intermediate data. This is especially beneficial for layers
with many tiles: for FN1.2, block-wise execution reduces the estimate
from 357.78 to 5.91 MiB and latency from 92.24 to 28.96 ms.
C can reduce storage further by using fewer tile groups and sharing
each active input block among channel workers. Additional channel
parallelism also requires more concurrent GEMM output buffers.
The balance between these effects gives C lower estimated storage
than B in four layers and the same storage in two. In the remaining
four layers, the larger estimate supports additional channel parallelism.
Across all ten layers, C's estimate remains at or below A's.

The waiting profiles show that most of A's barrier waiting occurs at
internal stage barriers. B removes these barriers, but final-barrier
waiting still reaches 93.75\% when tile parallelism is limited.
C's internal-barrier waiting ranges from 0.00\% to 4.24\%, and its
final-barrier waiting is lower than B's in nine of the ten layers.

The VN5.2 results also illustrate the trade-off between intermediate
storage and parallelism. B has only one active tile worker. C executes
more channel work concurrently, increasing the estimated intermediate
storage from 1.20 to 2.25 MiB and introducing 4.24\% internal-barrier
waiting. It reduces final-barrier waiting from 93.70\% to 0.66\% and
latency from 9.412 to 0.737 ms. Here, the additional active storage and
group synchronization support greater channel parallelism and reduce
idle waiting when tile parallelism is limited.

\noindent\textbf{\textit{Assembly vs. Intrinsics.}}
Figures~\ref{fig:fig_6a} and~\ref{fig:fig_6b} compare the single-thread runtimes of the hand-written AArch64 assembly and ARM NEON intrinsics implementations. The intrinsics implementation follows the proposed Winograd framework and uses the same execution schedule and parameter configuration as the assembly implementation. The two versions differ only in the low-level implementation of the transformation and GEMM micro-kernels. This comparison therefore isolates the performance effect of assembly-level optimization. Across the ten evaluated layers, assembly reduces the combined runtime of the input and output transformations by $11.0\%$ to $31.1\%$ and the GEMM runtime by $16.8\%$ to $35.5\%$.

For transformation, the performance difference mainly reflects register
allocation and instruction scheduling. With intrinsics, the compiler
determines whether reusable values remain in physical SIMD registers and
how strided loads, arithmetic instructions, and stores are
ordered~\cite{ref43, ref44}. The assembly kernels instead fix the register
mapping and interleave memory operations with transformation arithmetic,
which can reduce register moves, reloads, and dependency stalls.

GEMM places greater and more persistent pressure on the SIMD register
file. The $(\alpha,\eta)=(4,16)$ micro-kernel maintains 16 vector
accumulator registers, whereas the $(\alpha,\eta)=(7,8)$ micro-kernel
maintains 14, in addition to the registers allocated to the transformed
input and filter operands. The assembly implementation assigns these
operands and accumulators to fixed registers and uses the ping-pong
schedule to overlap operand loading with FMA execution. With intrinsics,
the compiler determines the final register assignment and instruction
order. Additional register moves, reloads, or exposed load latency recur
throughout the reduction over $C$, so their costs accumulate across the
GEMM computation.

Collectively, these ablation evaluations underscore the substantial efficiency improvements delivered by our proposed optimizations.

  \subsection{Single-core Performance Evaluation}
  \label{subsec:singlecore}

\begin{figure*}[t]
    \centering
    \includegraphics[scale=0.4]{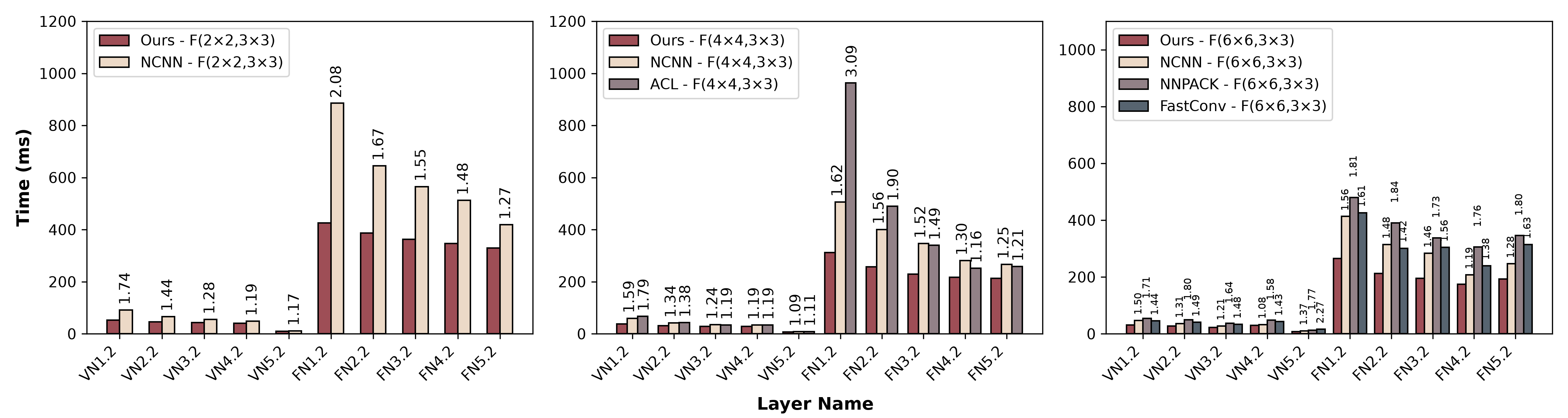}
    \caption{Step-wise comparison of the convolution layers' runtime against NCNN, NNPACK, FastConv and ACL with the same $F(m,r)$ on the Kunpeng920. 
    Each point on the x-axis represents a different layer, with VN and FN being abbreviations for VggNet and FusionNet, respectively.
    The y-axis denotes the runtime in milliseconds ($ms$). 
    The left figure shows $F(2\times2,3\times3)$, the middle figure shows $F(4\times4,3\times3)$ and the right figure shows $F(6\times6,3\times3)$.
    Each number above the bars represents the speedup our approach achieves compared to the corresponding library.
    }
    \label{fig:fig_7}
   \end{figure*}

  Moreover, to rigorously validate our method's performance benefits across various Winograd $F(m,r)$ configurations, 
  we conducted a stepwise evaluation comparing our approach against alternative libraries, 
  each evaluated using its corresponding $F(m,r)$ variant on the Kunpeng920. 
  For NCNN, we manually disabled its dynamic selection mechanism to ensure a fair baseline. 
  The results, presented in Figure~\ref{fig:fig_7}, 
  indicate that our method consistently outperforms the other libraries on a single-core basis for every layer of the chosen benchmark networks.
 
  Specifically, relative to NCNN, 
  our approach achieves up to $2.08\times$, $1.62\times$, and $1.56\times$ speedups for $F(2\times2,3\times3)$, $F(4\times4,3\times3)$, and $F(6\times6,3\times3)$, respectively. 
  Compared to ACL, NNPACK, and FastConv—with each employing its respective $F(m,r)$ variant—our method yields speedups ranging from $1.11\times$ to $3.09\times$, $1.58\times$ to $1.84\times$, and $1.38\times$ to $2.27\times$, respectively.

\begin{figure}[h!]
  \centering
  \includegraphics[scale=0.35]{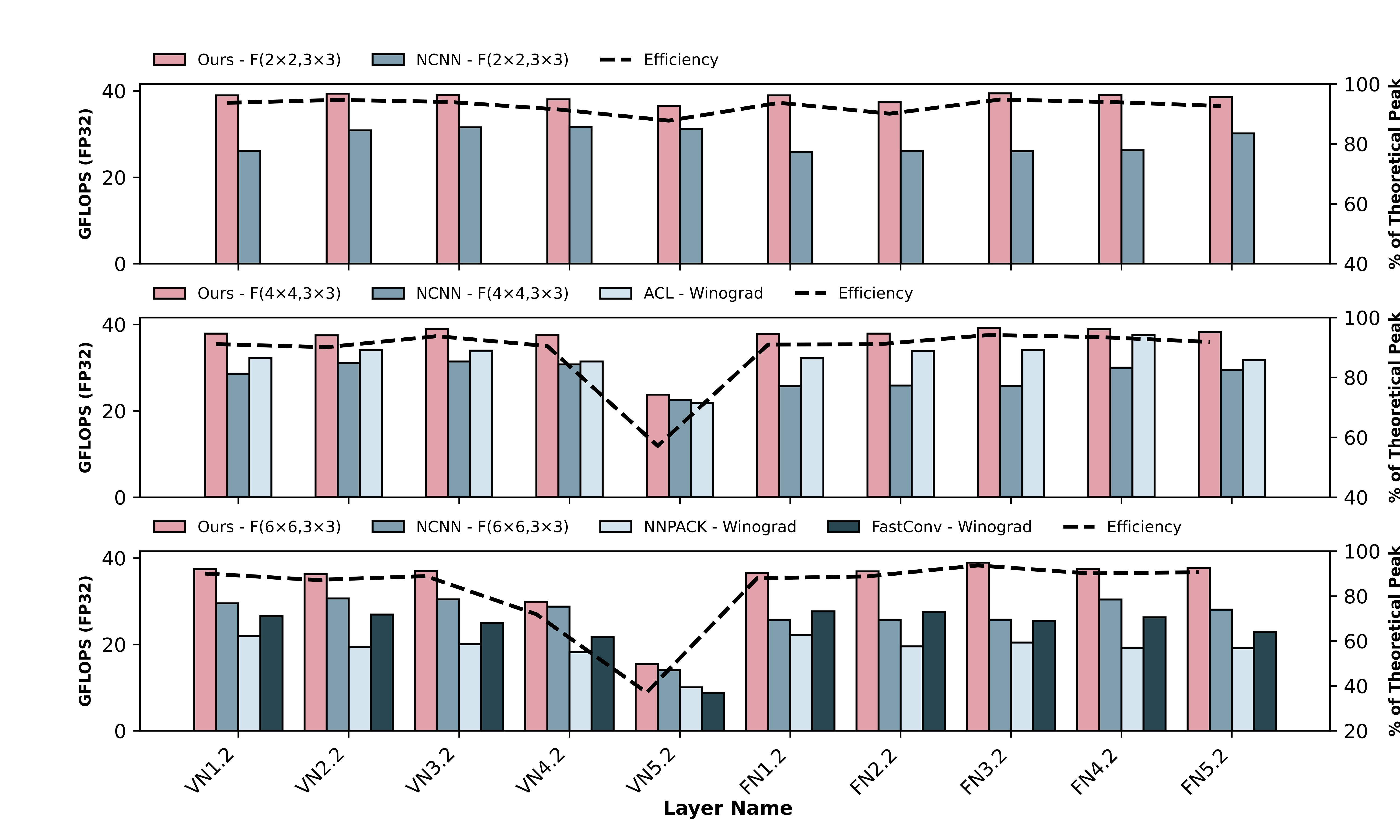}
  \caption{Layer-wise evaluations of the computational performance against NCNN, NNPACK, FastConv and ACL with the same $F(m,r)$.
  In the figure, the left y-axis represents performance in GFlop/s, 
  while the right y-axis indicates the fraction of the theoretical single-core peak performance (41.6 GFlop/s) on the Kunpeng 920. 
  For VggNet5.2, the small problem scale results in a memory-bound scenario, 
  leading to decreased GFlop/s for $F(4\times4,3\times3)$ and $F(6\times6,3\times3)$.
  }
  \label{fig:fig_8}
 \end{figure}

 Furthermore, we evaluated the computational efficiency of our approach in GFlop/s on the Kunpeng 920, 
 comparing it against NCNN, NNPACK, FastConv, and ACL under identical $F(m,r)$ configurations (see Fig.~\ref{fig:fig_8}). 
 As discussed in Equation~\eqref{eq:14}, increasing $L$ (where $L=16$ for $F(2\times2,3\times3)$, $L=36$ for $F(4\times4,3\times3)$, and $L=64$ for $F(6\times6,3\times3)$) intensifies data movement overhead. 
 This escalation leads to a downward trend in computational efficiency as $m$ increases when employing our GEMM-based implementation.
   
 For smaller layers, such as VggNet5.2, a pronounced decrease in GFlop/s is observed. 
 This is attributed to the limited computational load, resulting in a low CMR. 
 Consequently, as $m$ increases, these layers become memory-bound—particularly in the $F(6\times6,3\times3)$ case—thereby restricting full hardware utilization.

 NNPACK relies on TEWMM, a Level-1 BLAS operation, which inherently provides lower computational efficiency than GEMM-based routines (Level-3 BLAS), 
 since Level-3 BLAS operations more effectively amortize memory access costs. Although TensorGEMM (employed by FastConv) outperforms TEWMM, 
 it still exhibits suboptimal performance compared to libraries leveraging GEMM operations.

 In comparison, our method consistently surpasses NCNN and ACL, both of which are GEMM-based, 
 and demonstrates progressively stronger performance as the computational workload increases. 
 Notably, our approach attains up to $94.81\%$ of the Kunpeng 920's single core theoretical peak performance for $F(2\times2,3\times3)$ in layer FN3.2.
 Moreover, for most benchmarked layers, our GEMM routines maintain efficiencies exceeding $90\%$ of the single processor theoretical peak performance.

\begin{figure*}[t]
  \centering
  \includegraphics[scale=0.4]{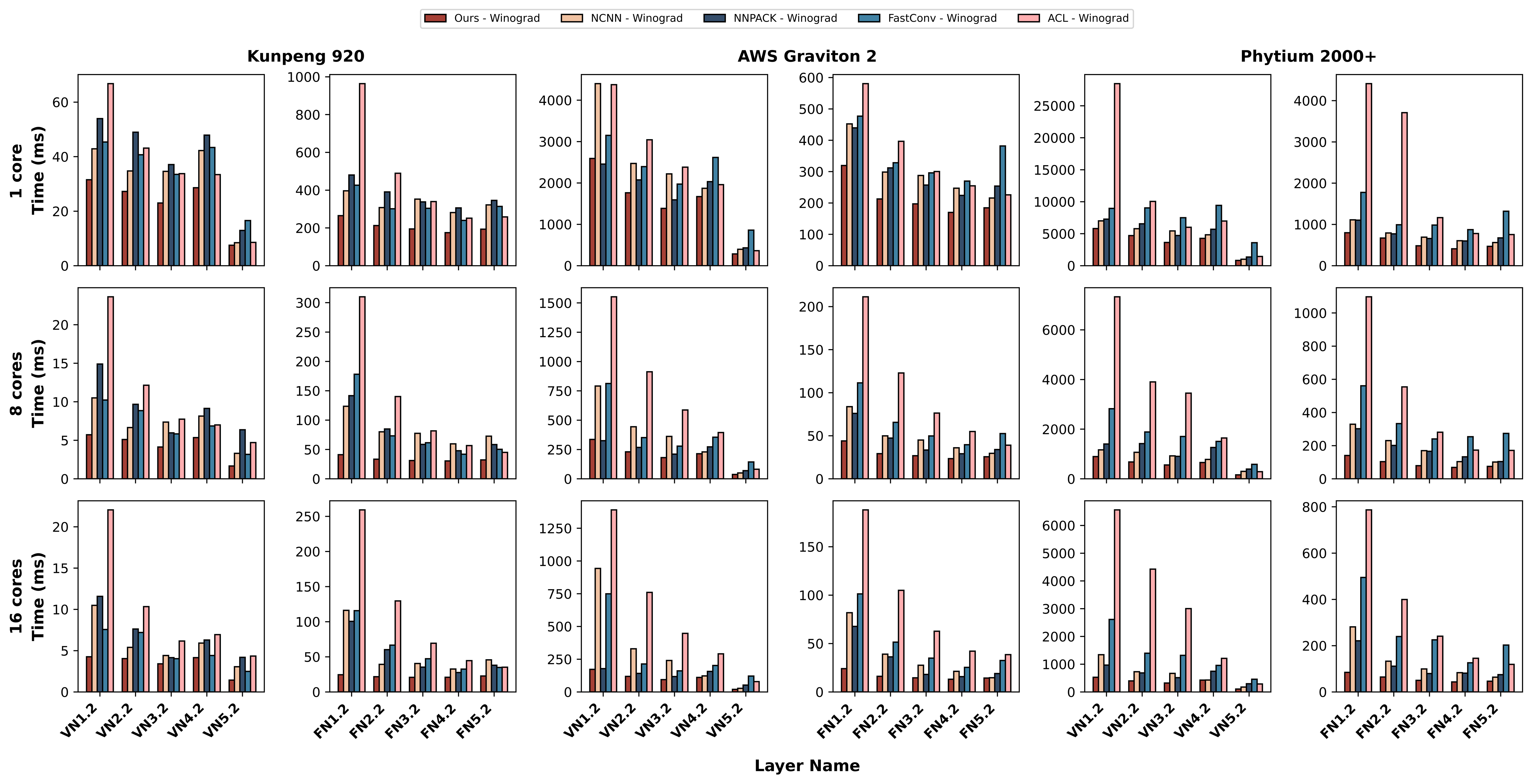}
  \caption{Step-wise comparison of convolution layer runtimes (single-core, 8-core, and 16-core) against NCNN, NNPACK, FastConv, and ACL on the Kunpeng 920, AWS Graviton2, and Phytium 2000+ platforms.
  }
 \label{fig:fig_9}
 \end{figure*}

 Next, we conducted a step-wise evaluation of our approach compared to other libraries across all benchmark layers, 
 as illustrated in Fig.~\ref{fig:fig_9}. 
 Our method employs a heuristic algorithm to determine the appropriate $F(m,r)$ configuration based on network scale, similar to NCNN, 
 and this selection process is performed at the initialization stage, thereby incurring no additional runtime overhead.
  All other libraries were evaluated with their default settings.

 On the Kunpeng 920 (single-core), our method achieves speedups of $1.13\times$ to $1.81\times$ over NCNN, 
 $1.61\times$ to $1.84\times$ over NNPACK, $1.37\times$ to $2.21\times$ over FastConv, and $1.14\times$ to $3.64\times$ over ACL. 
 On the AWS Graviton2 platform (single-core), the observed speedups range from $1.12\times$ to $1.70\times$ against NCNN, 
 $0.95\times$ to $1.52\times$ against NNPACK, $1.22\times$ to $3.02\times$ against FastConv, and $1.17\times$ to $1.86\times$ against ACL. 
 Similarly, on the Phytium 2000+ (single-core), our approach outperforms NCNN, NNPACK, FastConv, and ACL by $1.13\times$ to $1.50\times$, 
 $1.15\times$ to $1.65\times$, $1.48\times$ to $4.38\times$, and $1.61\times$ to $5.53\times$, respectively.

\subsection{Multi-core Performance Evaluation}
\noindent\textbf{\textit{Single-batch evaluation ($N=1$).}}

For the single-batch evaluation, we measure multi-core performance
using 8 and 16 threads on the Kunpeng 920, AWS Graviton2, and Phytium
2000+ platforms, as shown in Fig.~\ref{fig:fig_9}.

\textbf{Average speedups: }
\begin{itemize}
  \item \textbf{Kunpeng 920:} With 8 threads, our method achieves mean speedups of $2.05\times$, $2.27\times$, $1.95\times$, and $3.01\times$ over NCNN, NNPACK, FastConv, and ACL, respectively. 
  With 16 threads, the corresponding speedups are $2.07\times$, $2.18\times$, $2.07\times$, and $3.78\times$.
  \item \textbf{AWS Graviton2:} At 8 threads, the corresponding speedups are $1.54\times$, $1.34\times$, $2.13\times$, and $2.31\times$, 
  rising to $1.94\times$, $1.63\times$, $2.97\times$, and $2.86\times$ at 16 threads.
  \item \textbf{Phytium 2000+:} With 8 threads, our method achieves mean speedups of $1.71\times$, $1.91\times$, $3.24\times$, and $4.58\times$ over NCNN, NNPACK, FastConv, and ACL, respectively. 
  With 16 threads, the corresponding speedups are $1.93\times$, $1.76\times$, $4.14\times$, and $6.46\times$.
\end{itemize}

\textbf{Maximum speedups: }
\begin{itemize}
  \item \textbf{Kunpeng 920:} Up to $3.00\times$, $3.81\times$, $4.33\times$, and $7.54\times$ with 8 threads, and $4.74\times$, $4.10\times$, $4.72\times$, and $10.57\times$ with 16 threads over NCNN, NNPACK, FastConv, and ACL, respectively.
  \item \textbf{AWS Graviton2:} Maximum gains reach $2.16\times$, $1.73\times$, $3.83\times$, and $4.91\times$ with 8 threads, and $5.46\times$, $2.81\times$, $6.11\times$, and $8.05\times$ with 16 threads. The 16-thread maxima over NCNN and ACL occur on VN1.2, while the maximum over FastConv occurs on VN5.2.
  \item \textbf{Phytium 2000+:} We observe speedups of up to $2.40\times$, $3.68\times$, $3.99\times$, and $7.80\times$ with 8 threads, and $3.32\times$, $2.78\times$, $5.84\times$, and $9.28\times$ with 16 threads.
\end{itemize}
The larger performance gap over ACL on some layers is associated with its stage-wise execution. ACL stores transformed inputs and outputs in full-layer intermediate buffers between stages, increasing the memory footprint and potentially generating more cache and memory traffic, particularly for layers with large spatial dimensions. Its runtime stages also expose different parallel granularities, which can affect scalability. Our block-wise schedule consumes intermediate data earlier, reuses buffers across blocks, and coordinates cache blocking with thread decomposition. On FN1.2, our $F(6\times6,3\times3)$ variant additionally requires fewer Winograd-domain multiplications per output element than ACL's $F(4\times4,3\times3)$. The largest reported gap occurs on this layer with 16 threads on the Kunpeng 920: ACL takes 259.188 ms and our method takes 24.519 ms, yielding a speedup of $10.57\times$.

\begin{figure}[h]
  \centering
  \includegraphics[scale=0.3]{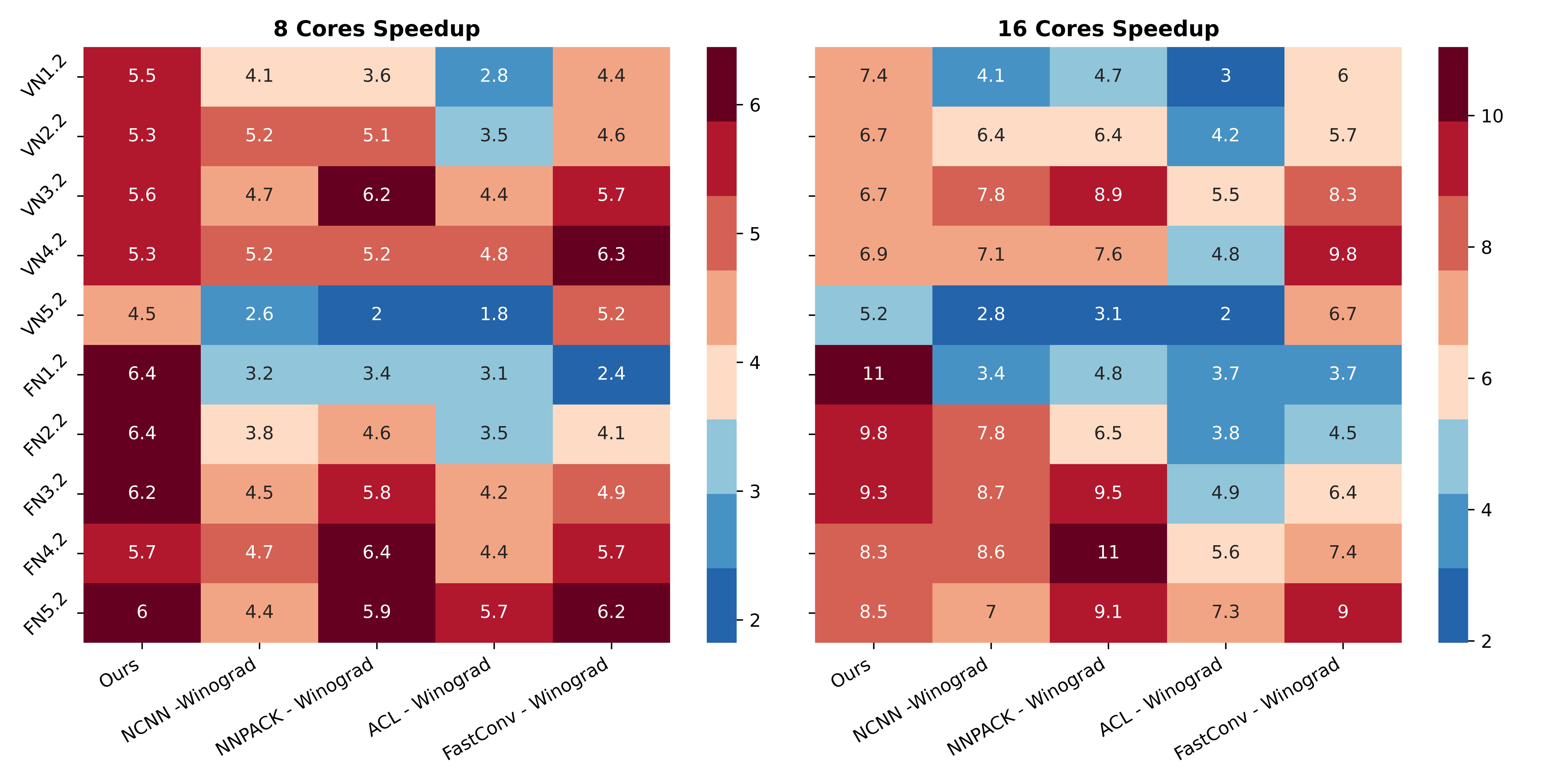}
  \caption{Layer-wise evaluations of the speedup of parallel execution against single-thread execution.
  }
  \label{fig:fig_10}
 \end{figure}

These results underscore the scalability and robust performance of our
approach across diverse platforms and threading configurations.
Figure~\ref{fig:fig_10} compares the parallel efficiency of our method
and the other libraries using 8 and 16 threads.
Our approach achieves higher parallel efficiency for most benchmark
layers, particularly those with larger workloads. For $N=1$,
inter-batch parallelism is unavailable. Layers with large spatial
dimensions generally provide more tiles, allowing threads to be
distributed primarily across the $T$ dimension. As the spatial
dimensions decrease and the channel counts increase, tile-level
parallelism becomes more limited, and the strategy shifts thread
decomposition toward the $C/K$ dimensions. This layer-shape-aware
decomposition helps maintain parallel efficiency across layers with
different tensor shapes.

As the thread count increases from 8 to 16, however, the available
parallelism represented by $T/T_{blk}$, $C/C_T$, and $K/K_{blk}$ is
insufficient in some layers to keep all cores equally occupied.
Meanwhile, the amount of computation assigned to each thread decreases,
making the intra-group synchronization after input transformation a
larger fraction of the total runtime. These effects reduce parallel
efficiency at higher thread counts and are most visible in VN5.2, whose
workload provides limited task-level parallelism.

\noindent\textbf{\textit{Multi-batch evaluation ($N=4,8,16$).}}

\begin{figure}[t]
  \centering
  \includegraphics[width=\columnwidth]{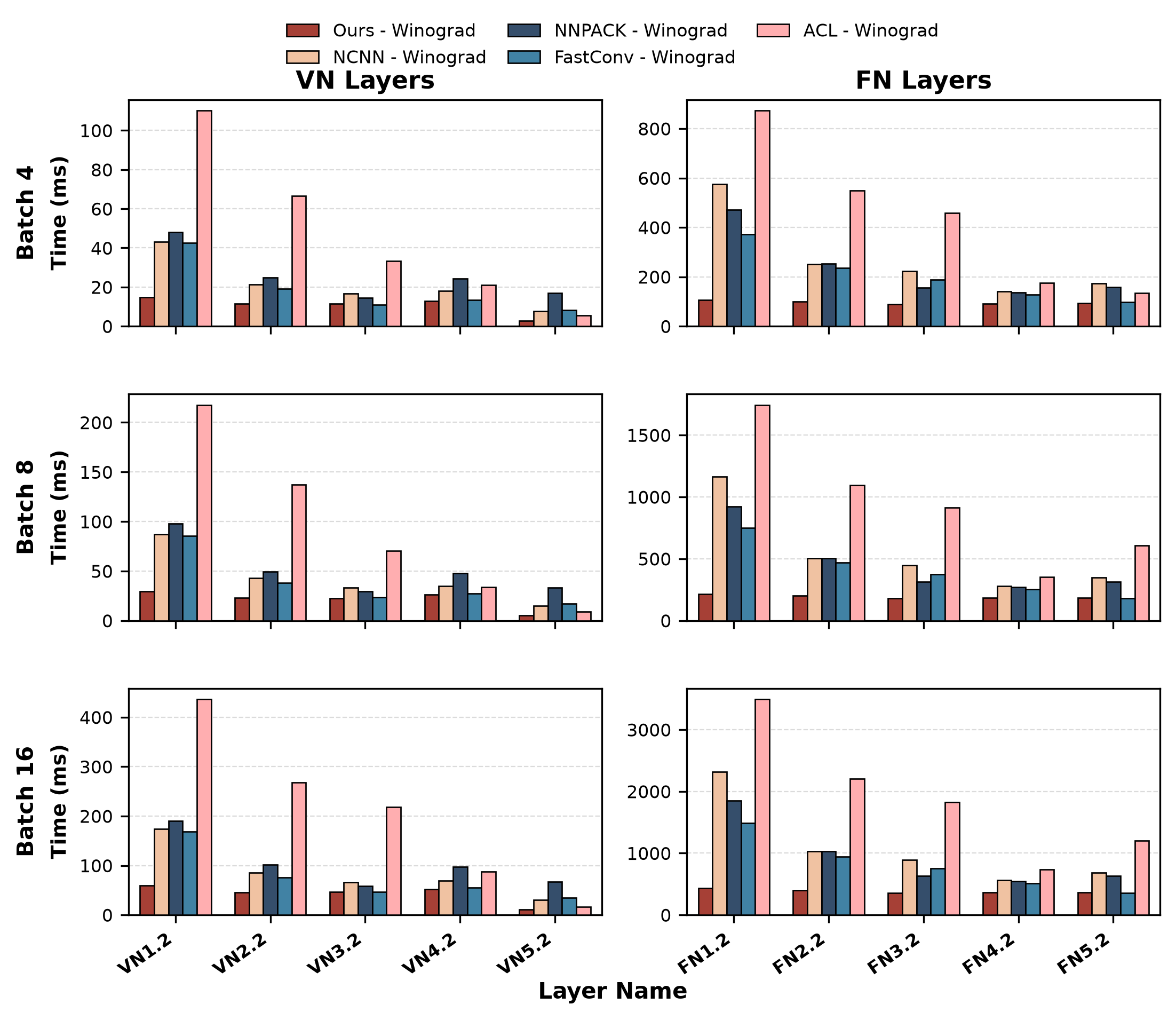}
  \caption{Layer-wise batch-latency comparison of Ours, NCNN, NNPACK, FastConv, and ACL performed on the Kunpeng 920 using 16 CPU cores at batch sizes 4, 8, and 16.}
  \label{fig:fig_11}
\end{figure}

We evaluate batch sizes of 4, 8, and 16 on the Kunpeng 920 using
16 CPU cores. Fig.~\ref{fig:fig_11} reports the batch latency of
the ten benchmark layers. Across these batch sizes, our method
achieves geometric mean speedups of 2.22 to 2.23$\times$ over NCNN,
2.35 to 2.36$\times$ over NNPACK, 1.80 to 1.82$\times$ over FastConv,
and 3.48 to 3.88$\times$ over ACL.

\begin{figure}[t]
  \centering
  \includegraphics[width=\columnwidth]{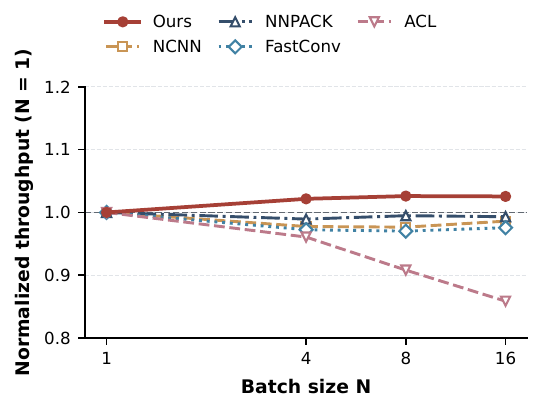}
  \caption{Normalized throughput across batch sizes on the Kunpeng 920 using 16 CPU cores.}
  \label{fig:batch_throughput}
\end{figure}

Fig.~\ref{fig:batch_throughput} shows the geometric mean normalized
throughput across the ten layers. For each implementation and layer,
normalized throughput is calculated as $R(N)=N\,t(1)/t(N)$, where
$t(N)$ is the median batch latency using 16 CPU cores. Our geometric
mean normalized throughput remains between 1.022 and 1.026 at
$N=4,8,16$, indicating that per-sample processing efficiency is
largely maintained relative to $N=1$.

 \subsection{Accuracy}
 Table \ref{tab:table2} reports the average and maximum errors across all layers of our implementation, as well as a comparison with other libraries.
The results use single-precision floating-point direct convolution as the reference baseline, 
with inputs and filters drawn from a uniform distribution within [-1.0, 1.0]. 
As shown, error magnitudes increase alongside larger $m$ values, reflecting the heightened sensitivity introduced by the transform matrices in Winograd convolution \cite{ref6}.

 Despite this trend, convolutional neural networks typically tolerate low-precision computations, 
 and errors below $10^{-2}$ have been reported to leave training and inference stability largely unaffected \cite{ref39, ref40}. 
 All $F(m,r)$ variants of our approach maintain error levels below this threshold, 
 thereby ensuring stable and reliable performance in practical applications.
 
 \begin{table*}[t]
  \centering
  \caption{Element errors of convolution neural networks.}
  \label{tab:table2}
  
  \begin{tabular}{c c|c c c c c c}
  \hline
  \multirow{2}{*}{Network} & \multirow{2}{*}{Error Type}  
  & \multirow{2}{*}{Ours-F($2^2, 3^2$)} 
  & \multirow{2}{*}{Ours-F($4^2, 3^2$)} 
  & \multirow{2}{*}{Ours-F($6^2, 3^2$)} 
  & \multirow{2}{*}{NCNN-F($2^2, 3^2$)} 
  & \multirow{2}{*}{NCNN-F($4^2, 3^2$)} 
  & {} \\ 
  & & & & & & & {} \\ 
  \hline
  \multirow{2}{*}{VggNet} & avg & 9.384078E-06 & 1.089130E-05 & 7.089612E-05 & 9.377054E-06 & 2.084713E-05 & {} \\
                         & max & 1.628480E-05 & 3.041010E-05 & 1.220090E-04 & 1.624360E-05 & 3.623090E-05 & {} \\
  \hline
  \multirow{2}{*}{FusionNet} & avg & 1.261121E-05 & 4.675881E-05 & 9.513018E-05 & 1.260211E-05 & 2.839505E-05 & {} \\
                             & max & 3.239750E-05 & 1.195620E-04 & 2.424290E-04 & 3.235360E-05 & 7.259690E-05 & {} \\
  \hline
  \end{tabular}
  
  \vspace{1em}
  
  \begin{tabular}{c c|c c c c c c}
  \hline
  {} & {} & \multirow{2}{*}{NCNN-F($6^2, 3^2$)} & \multirow{2}{*}{NNPACK} & \multirow{2}{*}{FastConv} & \multirow{2}{*}{ACL} & {} & {} \\ 
  {} & {} & & & & & {} & {} \\ 
  \hline
  \multirow{2}{*}{VggNet} & avg & 7.023788E-05 & 4.903992E-05 & 7.914980E-05 & 3.513735E-05 & {} & {} \\ 
                         & max & 1.220840E-04 & 7.423270E-05 & 1.383230E-04 & 6.083400E-05 & {} & {} \\
  \hline
  \multirow{2}{*}{FusionNet} & avg & 9.406444E-05 & 5.613214E-05 & 1.176832E-04 & 4.370908E-05 & {} & {} \\ 
                             & max & 2.379180E-04 & 1.085370E-04 & 2.993280E-04 & 1.191110E-04 & {} & {} \\
  \hline
  \end{tabular}
  
  \end{table*}

 \section{RELATED WORK}
 \label{sec:relatedwork}
 Since the pioneering work of Lavin and Gray \cite{ref6}, 
 who introduced the Winograd Minimal Filter Algorithm to reduce the computational complexity of convolutional operations in convolutional neural networks (CNNs), 
 significant efforts have been devoted to optimizing Winograd-based convolution methods across diverse architectures and optimization layers. 
 Jia et al.~\cite{ref30} developed an N-dimensional Winograd
 implementation for arbitrary kernel sizes on x86 many-core CPUs.
 Xie et al.~\cite{ref46} later presented HAWC, an FP16 Winograd
 implementation for ARM many-core processors that uses a stage-wise
 execution schedule, FP16 packing, a JIT-generated NEON GEMM kernel,
 and AArch64-specific memory instructions.
 Li et al.~\cite{ref31} improved performance on ARM many-core processors
 by integrating TEWMM and GEMM kernels and using a NUMA-aware scheduling
 strategy to mitigate remote memory accesses and cache contention.
 Gupta et al.~\cite{ref48} studied the co-design of CNN kernels and
 long-vector architectures for ARM-SVE and RISC-VV. Their ARM-SVE
 Winograd implementation used vector-length-agnostic inter-tile
 parallelism across channels to exploit long vectors in the
 transformation kernels.
 
 Further optimization paradigms have been explored in other works. 
 Lan et al. \cite{ref32} proposed FeatherCNN, which introduced the TensorGEMM subroutine to optimize Winograd convolution by reducing memory movement and improving register blocking efficiency.
 Building on FeatherCNN, Meng et al. \cite{ref14} proposed FastConv, which enhanced FeatherCNN by incorporating automatic kernel generation and auto-tuning strategies, 
 allowing it to achieve better performance portability and adaptability across various ARM CPU configurations. 
 For Intel Xeon Scalable Processor environments, 
 Wang et al. \cite{ref33,ref34} demonstrated enhancements to LoWino, 
 leveraging VNNI instructions to accelerate low-precision computations.
 Dolz et al.~\cite{ref41} presented a portable Winograd convolution
 implementation for multi-core CPUs using platform-specific SIMD
 intrinsics and OpenMP, with performance competitive with an
 im2col+GEMM baseline.
 
 Substantial progress has also been made on GPU platforms. 
 Yan et al. \cite{ref35} improved single-precision Winograd convolution performance on NVIDIA Volta and Turing architectures by conducting SASS-level optimizations and refining memory access patterns. 
 Jia et al. \cite{ref36}, in turn, introduced a MegaKernel-based fusion approach and a novel task mapping algorithm to reduce task dependency overheads and achieve better resource balance.
 
 Furthermore, advances have been made to extend the applicability of the Winograd algorithm to larger kernels and more complex convolutional settings.
  Huang et al. \cite{ref37} proposed the Decomposable Winograd Method (DWM), which decomposes large kernels into smaller ones to accommodate large kernel and stride scenarios. 
  Similarly, Yang et al. \cite{ref38} introduced the Stride-based Convolution Decomposition Method (SCDM), 
  enabling the application of Winograd, FFT, and FFA-based accelerations over a variety of convolution sizes and strides.
 
  In contrast to TEWMM-based~\cite{ref31} and TensorGEMM-based
  implementations~\cite{ref14,ref32}, Jia et al.~\cite{ref30},
  LoWino~\cite{ref34}, and our method formulate the Winograd-domain
  computation as a batch of independent matrix multiplications. Jia et
  al. scatter the transformed inputs and filters into separate matrices
  for each Winograd-domain coordinate and execute input transformation,
  Batched-GEMM, and output transformation as global stages. This
  stage-wise organization materializes transformed inputs and GEMM
  outputs across stage boundaries. LoWino also adopts the Batched-GEMM
  formulation, but targets low-precision convolution on x86 CPUs using a
  customized data layout, multidimensional cache blocking, and
  register-blocked matrix multiplication. Its extension~\cite{ref33}
  evaluates both fused and non-fused schedules.

  We retain the Batched-GEMM formulation but coordinate the runtime
  stages at block granularity. The transformation kernels write their
  results directly into a custom z-shaped, GEMM-ready layout in the
  order consumed by GEMM. Unlike the fused path of LoWino, which
  combines the three stages for a partition of tiles into a single
  kernel, our framework retains separate micro-kernels for input
  transformation, GEMM, and output transformation and invokes them
  within the same block-wise runtime loop. For ARMv8, this schedule and
  layout are co-designed with hand-tuned AArch64 transformation and GEMM
  micro-kernels. A performance model selects the blocking parameters
  under L1 and L2 cache-capacity constraints. The independent
  Winograd-domain matrices provide one source of parallelism, while our
  multi-core strategy further selects and combines parallelism across
  the batch, tile, input-channel, and output-channel loops according to
  the shape of each layer and coordinates this decomposition with the
  block-wise schedule.
 
  Prior work covers both algorithmic extensions of Winograd convolution
  and architecture-specific implementations for x86 and ARM multi-core
  CPUs, long-vector processors, and GPUs. These studies explore
  alternative Winograd-domain computation schemes, low-precision
  arithmetic, data-layout and cache optimizations, kernel generation,
  stage fusion, and parallel task scheduling. Collectively, these studies
  have extended the applicability of Winograd convolution and improved
  its execution efficiency across different processor architectures.
 
 \section{Conclusion and future work}
 \label{sec:conclusion}
 This work presents an ARMv8 Winograd convolution framework based on
 a block-wise runtime schedule. The schedule coordinates input
 transformation, GEMM, and output transformation and uses a performance
 model to select blocking parameters under L1 and L2 cache-capacity
 constraints. The transformation kernels write directly to a custom
 z-shaped, GEMM-ready layout, so GEMM requires no separate repacking
 step. The transformation and GEMM micro-kernels are hand-tuned in
 AArch64 assembly. A layer-shape-aware parallel strategy distributes
 threads across the batch, tile, input-channel, and output-channel
 loops.

 We evaluate the proposed method on ten convolution layers using three
 ARMv8 processors. Across these processors, the 16-thread mean speedups
 over NCNN, NNPACK, FastConv, and ACL range from $1.63\times$ to
 $6.46\times$.
 The GEMM micro-kernel reaches up to $94.81\%$ of the theoretical
 single-core peak on the Kunpeng 920. On the same processor, the
 multi-batch evaluation also shows the best overall performance at
 batch sizes 4, 8, and 16. Future work will extend the framework to
 3-D and larger-kernel Winograd convolution and port its micro-kernels
 to other vector ISAs.

\begin{IEEEbiographynophoto}{Haoyuan Gui}
is currently pursuing the Ph.D. degree with the Institute of Software,
Chinese Academy of Sciences, Beijing, China. His research interests
include high-performance computing, deep learning acceleration, and
performance optimization on modern processor architectures.
\end{IEEEbiographynophoto}

\begin{IEEEbiographynophoto}{Xiaoyu Zhang}
received the B.E. degree in computer science and technology from the
University of Science and Technology Beijing, Beijing, China, in 2023.
She is currently pursuing the M.E. degree in computer science and
technology with the Institute of Software, Chinese Academy of Sciences,
Beijing, China. Her research interests include high-performance
computing and parallel programming.
\end{IEEEbiographynophoto}

\begin{IEEEbiographynophoto}{Chong Zhang}
received the B.S. degree in software engineering from Dalian University
of Technology, Dalian, China, in 2019. He is currently pursuing the
Ph.D. degree with the Institute of Software, Chinese Academy of Sciences,
Beijing, China. His research interests include high-performance
computing and parallel algorithms.
\end{IEEEbiographynophoto}

\begin{IEEEbiographynophoto}{Huiyuan Li}
received the Ph.D. degree in computational mathematics from Shanghai
University, Shanghai, China, in 2001. He is currently a research professor
with the Institute of Software, Chinese Academy of Sciences, Beijing,
China, where he is also the executive deputy director of the Laboratory
of Parallel Software and Computational Science. His research interests
include high-performance computing software and algorithms, mathematical
libraries, spectral methods for partial differential equations, fast
transforms, and eigenvalue problems. Prof. Li is the secretary-general
of the Spectral Methods Committee of the China Society for Industrial
and Applied Mathematics.
\end{IEEEbiographynophoto}

\end{document}